\documentclass[twocolumn]{aastex701}
\usepackage{natbib,graphicx}
\usepackage{longtable}
\usepackage{booktabs}
\usepackage{makecell}
\usepackage{tabularx}
\usepackage{gensymb}
\usepackage{subcaption}
\usepackage{caption}
\usepackage{hyperref}
\usepackage{orcidlink}
\usepackage[export]{adjustbox}
\graphicspath{ {./s_n_pics/} }
\begin{document}
\title{On the Use of Cyclic Spectroscopy for Canonical Pulsars}
\author{Lawrence Wise, \orcidlink{0009-0005-8421-128X}}
\affiliation{\rm{Department of Astronomy, University of Virginia, Charlottesville, VA 22904, USA}}
\affiliation{\rm{Green Bank Observatory, P.O. Box 2, Green Bank, WV 24944, USA}}
\email{evc9pu@virginia.edu}
\author{Jacob Turner, \orcidlink{0000-0002-2451-7288}}
\thanks{Jansky Fellow of the National Radio Astronomy Observatory}
\affiliation{\rm{Green Bank Observatory, P.O. Box 2, Green Bank, WV 24944, USA}}
\affiliation{\rm{Department of Astronomy and Cornell Center for Astrophysics and Planetary Science, Cornell University, Ithaca, NY 14850, USA}}

\email{jeturner@nrao.edu}
\date{June 2026}
\begin{abstract}
We use the cyclic figure of merit to determine the likelihood of achieving cyclic deconvolution for 2609 canonical pulsars, a population for which this technique has not previously been used. These pulsars have spin periods between 0.1 and 10 seconds and are surveyed across 15 different telescope-observing frequency combinations. We conclude that cyclic deconvolution will only be a viable method for pulse broadening function or phase retrieval recovery in a handful of canonical pulsars. Few, if any, pulsars pass the conservative threshold for a given telescope-frequency combination. We believe that the cyclic figure of merit does not accurately reflect our ability to use cyclic deconvolution on this population, instead showing a shortcoming of the measurement when observing canonicals at low observing frequencies. Still, it is reliable for MSPs and for canonicals at high observing frequencies. More promising is the prospect of using this technique to recover scintles for canonical pulsars; DSA and FAST are best-suited for this as judged by the number of sources where scintles may be resolved and S/N is sufficient. We suspect there may be more sources that would have large figures of cyclic merit, but insufficient S/N, at very low frequencies. We advocate for expansions of real-time cyclic spectroscopy backends for current and upcoming low frequency telescopes. 
\end{abstract}

\section{Introduction}
\label{sec:intro}
Cyclic spectroscopy is a signal processing technique that can utilize the periodicity and harmonic information of pulsar emission to extract more information from data than is available through phase-resolved spectroscopy \citep{demorest2011}. The cyclic spectrum for pulsar emission interacting with an ionized medium, such as the interstellar medium (ISM), is given by
\begin{equation}
    S_E(\nu, \alpha_k) = (H(\nu + \alpha_k/2)H^*(\nu - \alpha_k/2))S_x(\nu,\alpha_k).
\end{equation}

Here, $\nu$ is the observing frequency, $\alpha_k$ is the cyclic frequency (meaning $\alpha_k = k/P$, where $k$ is an integer and $P$ is the pulse period), $S_x(\nu, \alpha)$ is the Fourier transform of the intrinsic pulse profile, and $H(\nu)$ is the transfer function of the ISM. The cyclic spectrum is valid up to the pulsar’s scintillation timescale. After this timescale is breached, the pulse signal must be resampled \citep{Walker_2013}. 

The pulse broadening function is the Fourier transform of the pulsar's transfer function, and it is used to measure delays caused by interstellar scattering. If the intrinsic pulse profile is deconvolved from the observed signal, one can use the recovered pulse broadening function/transfer function to quantify aspects of the pulsar’s time-of-arrival (TOA) delays that have been historically difficult to measure, stack the recovered transfer functions in time to create a dynamic wavefield for an observation, or even circumvent the Gabor limit to achieve high pulse phase resolution and frequency resolutions down to $1/P$ at the same time. Accurate TOA delay values allow us to improve pulsar timing array sensitivity to gravitational waves \citep{Palliyaguru_2015}. The scattering delays recovered by this method are considered to be more precise and accurate when compared to indirect inference \citep{Turner_2023}. A dynamic wavefield can show the full effect of ISM-induced spreading on the pulsar’s spectrum as the emission propagates, allowing us to reconstruct the time and frequency-evolving signal \citep{Walker_2013, Turner_2025}. Without the Gabor Limit, we can both study the ISM and precisely time the pulsar. This can give a clearer picture of the structure of the ISM along the line of sight. 

Cyclic merit is the measure of the likelihood that deconvolution will be possible on a given source for a given frequency, telescope, and backend receiver. Deconvolution is one of the main goals of cyclic spectroscopy, along with scintle resolution, so having a shorthand to calculate its viability can be invaluable. We calculated the cyclic figure of merit as
\begin{equation}
    m_{\rm{cyc}} \approx 2\pi(\textrm{S/N}) \frac{\tau_d}{\sqrt{PW_e}},
\end{equation}
which is an approximation based on the mathematics in \cite{Turner_2026}. Here, $\tau_{d}$ is the scattering delay, $P$ is the period, and $W_e$ is equivalent pulse width (calculated from the full-width half-max of the pulse). S/N is given by 

\begin{equation}
    \textrm{S/N} = \frac{SG}{T}\sqrt{2\Delta t_d \Delta \nu_d} \sqrt{\frac{P - W_e}{W_e}}.
\end{equation}
Here, $S$ is the source flux density, $G$ is the instrument gain, $T$ is the system temperature, and $\Delta t_d$ is the scintillation timescale. We used \cite{Turner_2026}'s calculation for $m_{\rm{cyc}}$ 2.0 because it is more accurate than the original $m_{\rm{cyc}}$. A similar update was mentioned in $m_{\rm{cyc}}$ in \cite{2021ApJ...913...98D}, appendix A. 

Cyclic spectroscopy has been performed on millisecond pulsars (MSPs) since the technique was first introduced into astronomy, but it has only been briefly tested on canonical pulsars. \cite{Turner_2026} showed that cyclic deconvolution may be possible on canonical pulsars, but a full survey has not been performed. The main benefit of cyclically deconvolving canonical pulsars is that they may have recoverable scintles through cyclic spectroscopy. The cyclic figure of merit is inversely proportional to the period and pulse width, two quantities which are often larger for canonicals. However, canonical pulsars may have higher signal-to-noise ratios and are often brighter than MSPs. Cyclic spectroscopy has not been performed on canonicals previously because their pulse profiles tend to be wider than those of MSPs, potentially making cyclic deconvolution difficult.  

Section \ref{sec:data} of the paper will discuss the important background information for this study, while Section \ref{sec:results} will discuss the prospects of each telescope-frequency combination.
Section \ref{sec:conclusions} will explain our findings and suggest future work. 

\section{Data}
\label{sec:data}
Our source selection consisted of using \textsc{psrqpy} \citep{Pitkin_2018} to search for all pulsars in the Australian Telescope National Facility (ATNF) pulsar catalogue \citep{Manchester_2005} with pulse periods less than ten seconds and greater than 0.1 seconds, and pulse widths measured via $W_{50}$. We also used several papers \citep{Johnston_1998, Bhat_1998, Main_2017, Stinebring_2022, Wu_2022, Liu_2022, Turner_2024_multiepoch} that had measured scattering delay, scintillation bandwidth, and scintillation timescale values. We pulled values for the $W_{10}$, $W_{50}$, and periods of these pulsars from the ATNF pulsar catalogue, as well as other relevant parameters. This resulted in 2609 pulsars. We calculated the system temperature by summing the telescope antenna temperature and the frequency- and location-dependent galactic background temperature, sourced from \textsc{pygdsm} \citep{2016ascl.soft03013P}. We used the \textsc{pulsar-spectra} package \citep{Swainston_2022} to extrapolate flux densities for those pulsars that did not have a measured flux density at all of our observing frequencies, but did have a measured flux density for some frequencies. The package requires between one and five points of data, depending on which fit it determines is best for the given data. The papers used by \textsc{pulsar-spectra} to inform the fits we used to infer our sources' flux densities were 
\cite{1973MNRAS.165..133M,1978A&A....68..361B,1978MNRAS.185..409M,1981Ap&SS..78...45I,1985ApJ...294L..25D,1985Natur.317..787S,1986ApJ...311..694S,1986AuJPh..39..103S,1991MNRAS.249..654M,1992MNRAS.255..401J,1993A&A...272L..13W,1993AstL...19..138M,1995A&AS..111..205S,1995ApJ...439..933L,1995ApJ...445..756C,1995MNRAS.273..411L,1996ApJ...456..305Z,1996ApJ...469..819C,1996MNRAS.279.1235M,1996MNRAS.282..691B,1997A&AS..126..121V,1997ApJ...474..426S,1997ApJ...485..820K,1997ApJ...488..364K,1997MNRAS.287..307V,1998ApJ...509..785S,1999A&AS..136..571H,2000A&AS..145..243K,1999ApJS..121..171W,2000ApJ...545.1007L,2000ARep...44..436M,2001AJ....121.1192M,2001AJ....121.3133G,2001AJ....122.2001C,2001MNRAS.328...17M,2002MNRAS.335..275M,2003MNRAS.342.1299K,2004A&A...425..949E,2004ApJ...600..905L,2004MNRAS.352.1439H,2005MNRAS.359..481K,2005MNRAS.359.1524L,2005MNRAS.363..929C,2005PhDT.......282C,2006MNRAS.369.1916J,2006MNRAS.372..777L,2007A&A...462..699K,2007AJ....134.1231C,2009A&A...498..223J,2009ApJ...701.1243D,2009MNRAS.398..943J,2011A&A...531A..16K,2011MNRAS.411.1575B,2011MNRAS.416..346K,2012ApJ...745..109L,2013ApJ...763...80B,2013ApJ...763...81L,2013MNRAS.431.3624Z,2014ApJ...791...67S,2014MNRAS.445.3105D,2015ApJ...808..156S,2015ApJ...812...81L,2015MNRAS.449.1869D,2015MNRAS.450.2922N,2016A&A...591A.134B,2016A&A...593A..21M,2016ApJ...817..130B,2016ApJ...829..119F,2016MNRAS.458.2509B,2016MNRAS.461..908B,2016RAA....16..159H,2017ApJ...840..108K,2017ApJ...845..156Z,2017PASA...34...20M,2017PASA...34...70X,2018MNRAS.473.4436J,2018MNRAS.474.2012B,2018MNRAS.474.4629J,2018MNRAS.475.1469B,2018MNRAS.479.2193R,2019A&A...626A.104S,2019ApJ...870....8S,2019ApJ...874...64Z,2019ApJ...875...19A,2019MNRAS.484.3691J,2019MNRAS.487.4332T,2019RAA....19..103X,2020A&A...635A..75B,2020A&A...635A..76B,2020ApJ...892...76M,2020MNRAS.491..725M,2020MNRAS.492.5878T,2021ApJ...923..211K,2021MNRAS.502.1253J,2021RAA....21..107H,2022ApJ...924..135P,2022MNRAS.512.4324K,2022PASA...39...42L,2023ApJ...956...40F,2023PASA...40...20B,2024ApJ...961...48W,2024MNRAS.530.1581K,2024ApJS..271...23D,2025ApJ...982..132K}.  If there was not enough data on a pulsar to use \textsc{pulsar-spectra}, we used a power law with a spectral index of -1.7 to estimate flux density. We used the \textsc{NE2025} electron density model to infer scintillation bandwidth, scintillation timescale, and scattering delay for all sources that did not have these values recorded \citep{Ocker_2026}. A transverse velocity of 100 km/s was assumed for all pulsars that did not have a measured transverse velocity in the ATNF catalogue.

If a given source had a measured $\Delta\nu_d$ or $\tau_d$, we can relate the two quantities via
\begin{equation}
    2\pi\tau_d\Delta\nu_d = C_1.
\end{equation}

\noindent $C_1$ is a constant that depends on the geometry of the ISM along the line of sight. We used 0.957 for $C_1$ \citep{Cordes_1998}, an estimation  that assumes the ISM can be described by thin-screen geometry and Kolmogorov turbulence. Since this estimation is used for calculating the scintillation bandwidth/scattering delay, as mentioned above, the derived quantity may be somewhat inaccurate depending on how well our chosen $C_1$ describes a given line of sight's turbulence and geometry. 

If uncertainties were not listed in ATNF or the aforementioned papers, errors of 50\% were assumed for scintillation bandwidths, timescales, scattering delays, and flux densities. If we had access to both scintillation bandwidth and scattering delay, they were used as-is, otherwise the scintillation bandwidth was estimated with NE2025 and the scattering delay was calculated from that value. We used an index of 4.4 for scaling scintillation bandwidth values to relevant observing frequencies. The index for scattering delay scaling was $-$4.4, and the index for scintillation timescale was 1.2. 

Cyclic merit was calculated as described in Section \ref{sec:intro}. We also came up with deconvolution regime boundaries, representing the edges of the regions wherein a signal may successfully be deconvolved; the upper boundary was calculated as 
\begin{equation}
    D_u = \frac{\nu}{ W_{10}^{1/4.4} \Delta\nu ^{1/4.4}}.
\end{equation}
This represents when in observing frequency the scattering tail becomes visible in the pulse profile, where the scintillation bandwidth is less than the inverse of the intrinsic pulse width.
The lower boundary was calculated as 
\begin{equation}
    D_{l} = \frac{\nu}{ P^{1/4.4} \Delta \nu ^{1/4.4}},
\end{equation}
as the cyclic spectrum can only achieve frequency channelizations equal to the pulsar's spin frequency or less before becoming undersampled in cyclic frequency \citep{Turner_2026}.
We have access to pulse period $P$ for nearly all pulsars through the ATNF catalogue, so we have more lower boundary calculations than upper. The upper boundary relies on $W_{10}$, a value that we are missing for several pulsars. However, the upper and lower boundaries both constitute a fairly even Gaussian distribution in log space as expected, so the lower number of upper deconvolution boundaries does not pose a significant detriment to our results. 

\section{Results and Discussion}
\label{sec:results}

\begin{deluxetable*}{CCCCCC}
\tablefontsize{\footnotesize}
\small
\setlength{\tabcolsep}{2pt}
\tablecolumns{6}

\tablecaption{Telescope Parameters\label{telescope}}
\tablehead{ \colhead{Telescope} &
\colhead{Frequency (MHz)} & \colhead{Receiver Temperature (K)} & \colhead{Gain (K/Jy)} & \colhead{Minimum Declination (\degree)} &
\colhead{Maximum Declination (\degree)}}

\startdata
\text{GBT}&350 &25 &2 &-45 &90 \\
\text{DSA}&800 &11&12.5 &-40 &90 \\
\text{MeerKAT}&600 &33&2.8 &-90 &42 \\
\text{FAST}&550 &33&13.67 & -14 &65.6 \\
\text{MWA}&150 &50& 1 &-90 & 25 \\
\text{MWA}& 200 &50&1 & -90 & 25 \\
\text{LWA}&74 &300&1.67 & -29 & 90 \\
\text{LOFAR} &150 &400&8.90 &0 & 90 \\
\text{LOFAR} &200 &400&5.01 & 0 & 90 \\
\text{NenuFAR}&48 &470.6& 8.60 &-23 &90 \\
\text{NenuFAR}& 85 &3095.0& 2.74 & -23 & 90 \\
\text{CHIME}& 450 &20& 1.16 & -11 & 90 \\
\text{uGMRT} & 150 &295& 8.58 & -53 & 90 \\
\text{uGMRT}& 200 &120& 8.58 & -53 & 90 \\
\text{uGMRT} & 350 &50& 8.58 & -53 & 90 \\
\enddata
\tablecomments{Parameters for each Telescope/Frequency combination used in the survey.}
\end{deluxetable*}

\noindent

\begin{deluxetable*}{CCCCCC}[htbp]
\tablefontsize{\footnotesize}
\small
\setlength{\tabcolsep}{2pt}
\tablecolumns{6}

\tablecaption{General Survey Results \label{table_gen_results}}
\tablehead{ \colhead{Telescope} &
\colhead{Total Visible Sources} & \colhead{Outside Full Decon. Regime} & 
\colhead{In Full Decon. Regime}&
\colhead{Outside Full Decon. Regime} &
\colhead{In Full Decon. Regime} 
\vspace{-.3cm}
\\  \colhead{} & \colhead{\text{with Calculated $m_{\rm cyc}$}} & \colhead{\text{\& Passing Optimistic}} & \colhead{\text{\& Passing Optimistic}}& \colhead{\text{\& Passing Conservative}} & \colhead{\text{\& Passing Conservative}}  \vspace{-.3cm}
\\  \colhead{} & \colhead{Out of 2609 Surveyed} & \colhead{\text{Threshold}} & \colhead{\text{Threshold}}& \colhead{\text{Threshold}} & \colhead{\text{Threshold}} }

\startdata
\text{FAST}$_{550}$& 1234 & 47 & 3 & 41 & 1 \\
\text{GBT}$_{350}$& 1879 & 37 & 1 & 35 & 0 \\
\text{DSA}$_{800}$& 1764 & 67 & 13 & 45 & 3 \\
\text{LWA}$_{74}$& 1560 & 33 & 11 & 33 & 11 \\
\text{LOFAR}$_{150}$& 951 & 34 & 5 & 33 & 4 \\
\text{LOFAR}$_{200}$& 951 & 33 & 2 & 33 & 2 \\
\text{NenuFAR}$_{48}$& 1443 & 35 & 13 & 33 & 12 \\
\text{NenuFAR}$_{85}$ & 1443 & 32 & 5 & 31 & 5 \\
\text{CHIME}$_{450}$ & 1220 & 34 & 0 & 33 & 0 \\
\text{uGMRT}$_{150}$ & 2061 & 47 & 7 & 37 & 6 \\
\text{uGMRT}$_{200}$ & 2061 & 53 & 4 & 40 & 3 \\
\text{uGMRT}$_{350}$& 2061 & 60 & 4 & 42 & 1 \\
\text{MeerKAT}$_{600}$ & 2330 & 39 & 1 & 35 & 0 \\
\text{MWA}$_{150}$ & 2168 & 33 & 5 & 33 & 5 \\
\text{MWA}$_{200}$ & 2168 & 34 & 2 & 33 & 2 \\
\enddata
\tablecomments{Overview of results for each Telescope/Frequency combination, including visible sources (based on source declination), inside and outside full deconvolution regime, and surpassing of the optimistic and/or conservative thresholds.}
\end{deluxetable*}

\begin{deluxetable*}{CCCC}
\tablecaption{Detectable and Resolvable Scintles\label{scintles}}
\tablehead{ \colhead{Telescope} & \colhead{Total} & \colhead{Telescope} & \colhead{Total}}
\startdata
\text{FAST}$_{550}$ & 193 & \text{GBT}$_{350}$ & 88 \\
\text{DSA}$_{800}$ & 399 & \text{LWA}$_{74}$ & 1 \\
\text{LOFAR}$_{150}$ & 8 & \text{LOFAR}$_{200}$ & 11 \\
\text{NenuFAR}$_{48}$ & 0 & \text{NenuFAR}$_{85}$ & 0 \\
\text{CHIME}$_{450}$ & 53 & \text{uGMRT}$_{150}$ & 17 \\
\text{uGMRT}$_{200}$ & 42 & \text{uGMRT}$_{350}$ & 144 \\
\text{MeerKAT}$_{600}$ & 204 &\text{MWA}$_{150}$ & 3 \\
\text{MWA}$_{200}$ & 5 &  &  \\
\enddata
\tablecomments{Number of scintles that are both detectable and resolvable for each telescope-frequency combination.}
\end{deluxetable*}

\begin{figure}
    \centering
    \includegraphics[width=0.5\textwidth]{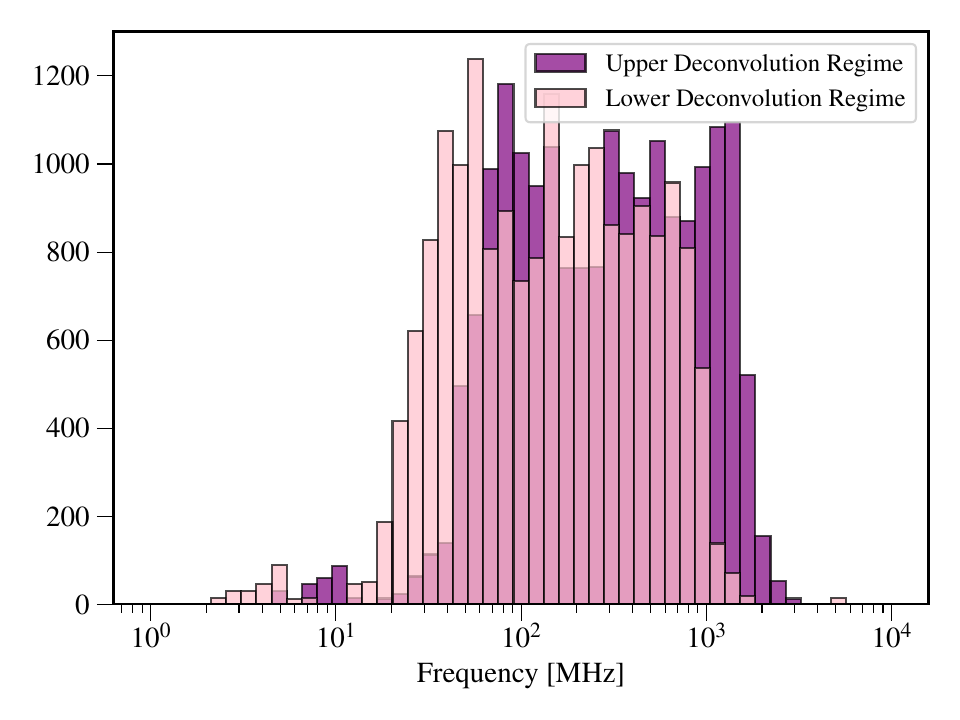}
    \caption{Histograms showing the lower (pink) and upper (purple) boundaries to the full deconvolution regime for all pulsars in our survey. The lower frequency boundary has more instances due to the upper frequency boundary requiring $W_{10}$ measurements, which were not available for all pulsars.}
\end{figure}

\subsection{GBT 350 MHz}
\begin{figure}[h]
\begin{minipage}{.45\textwidth}
    \centering
    \includegraphics[width=0.9\textwidth,left]{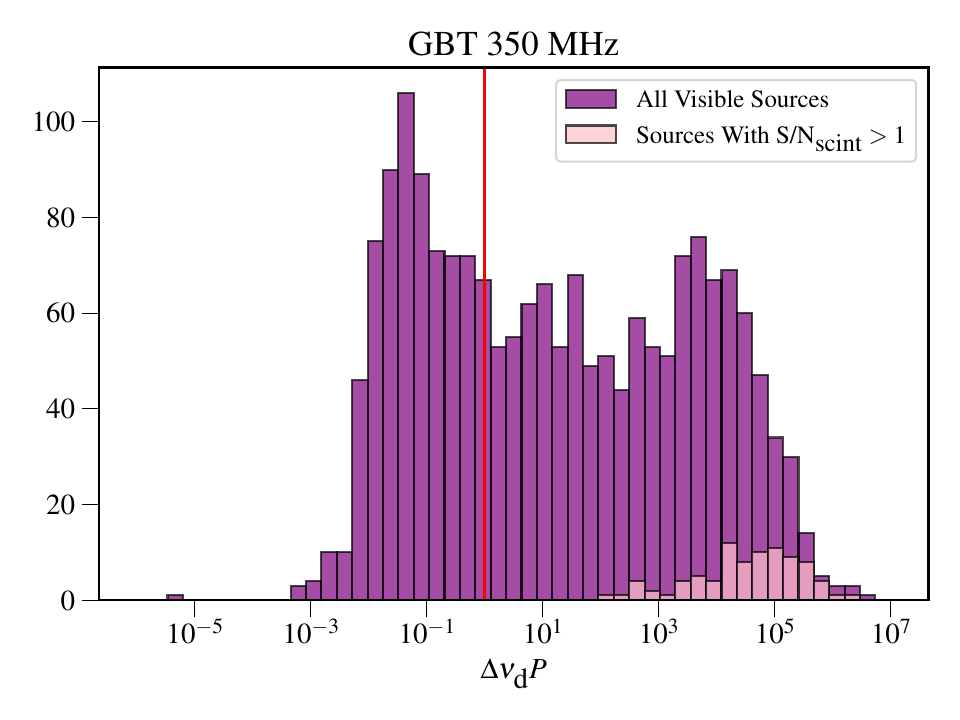}
    \caption{Expected scintle resolvability for all sources analyzed for GBT at 350 MHz (purple), and the subset of sources with S/N$>$1 (pink). We resolve scintles when $\Delta\nu_{\rm d}P>1$, as indicated the by the solid red vertical line. There could be some sources that pass the threshold within the margin of error, and thus do not show up in the histogram despite technically qualifying. There also could be some sources that would achieve resolution if observed at slightly higher frequency.}
    \label{fig:gbt_scintle}
\end{minipage}
\hspace{1cm}
\begin{minipage}{.45\textwidth}
    \centering
    \includegraphics[width=0.9\textwidth, left]{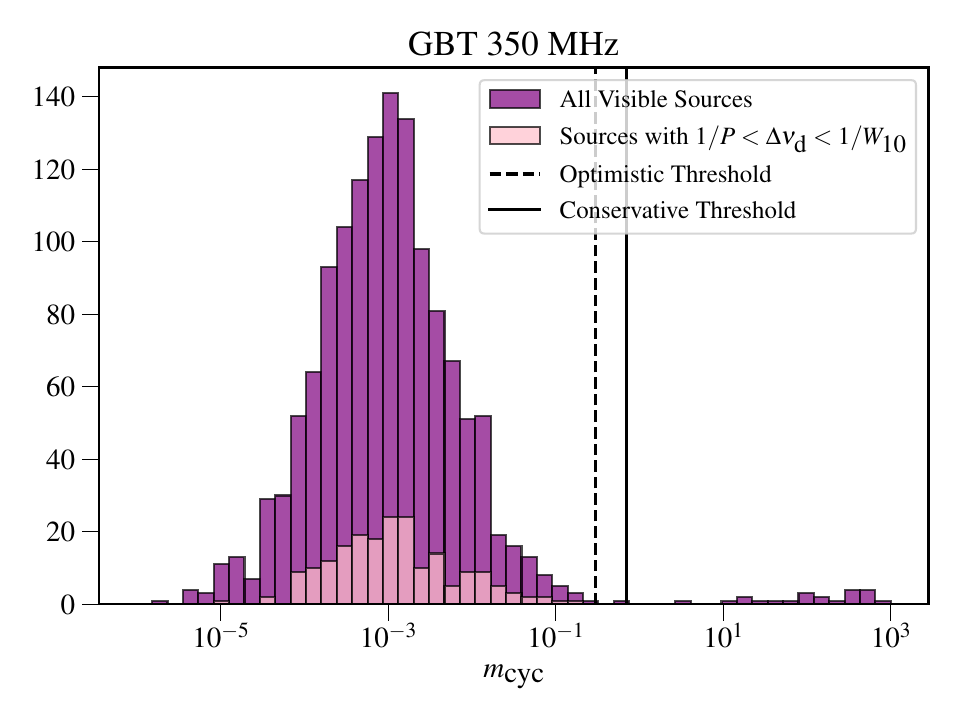}
    \caption{Cyclic merits for all sources analyzed for GBT at 350 MHz (purple), and the subset of sources in the full deconvolution regime (pink). The dashed line shows the optimistic cyclic merit threshold, while the dotted black line shows the conservative cyclic merit threshold. The same error issues discussed in the Figure 2 caption may also apply here.}
    \label{fig:gbt_cyclic}
\end{minipage}
\end{figure}

The GBT has the world’s first cyclic spectroscopy backend.  The GBT is not altogether promising for cyclic deconvolution on canonical pulsars,  with one source (J1935+1616) passing the optimistic threshold and no sources passing the conservative threshold, if only considering those in the full deconvolution regime. However, 88 sources have resolvable scintles. This indicates that the GBT would still be a useful instrument for studying the ISM with canonical pulsars using cyclic spectroscopy. An ultrawideband receiver from 0.7 to 4 GHz has been installed, which provides an increased frequency coverage for this instrument. The scintle resolvability is shown in Figure \ref{fig:gbt_scintle} and the histogram for all sources' cyclic figures of merit is Figure \ref{fig:gbt_cyclic}. 

\subsection{DSA 800 MHz}
\begin{figure}[h]
\begin{minipage}{.45\textwidth}
    \centering
    \includegraphics[width=0.9\textwidth, right]{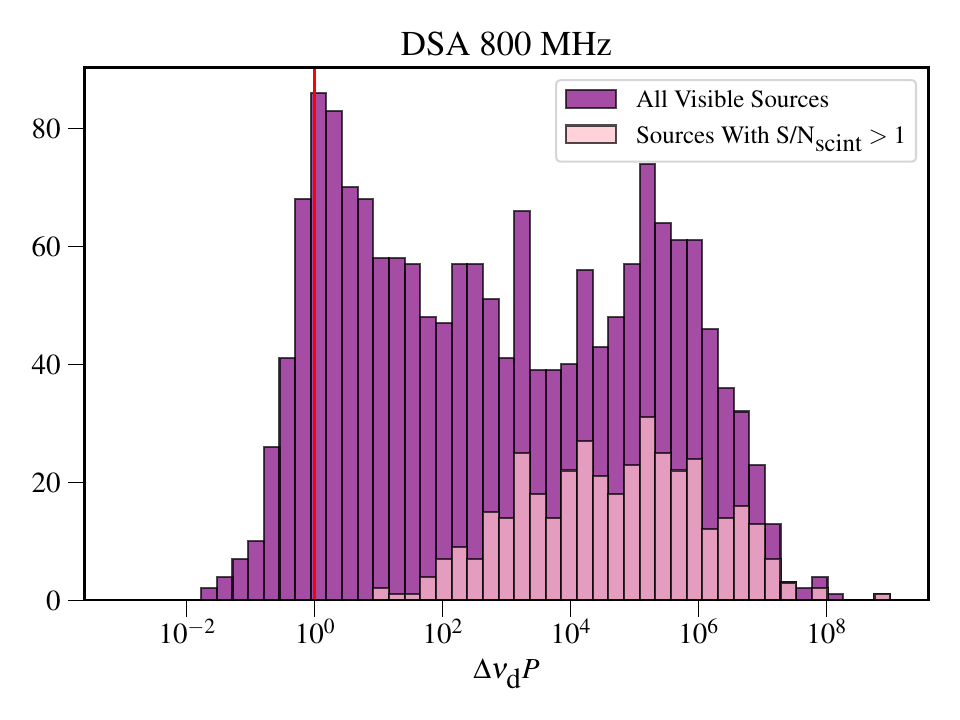}
    \caption{Expected scintle resolvability for all sources analyzed for DSA at 800 MHz (purple), and the subset of sources with S/N$>$1 (pink). Plot description same as that of Figure 2.}
    \label{fig:dsa_scintle}
\end{minipage}
\hspace{1cm}
\begin{minipage}{.45\textwidth}
    \centering
    \includegraphics[width=0.9\textwidth, right]{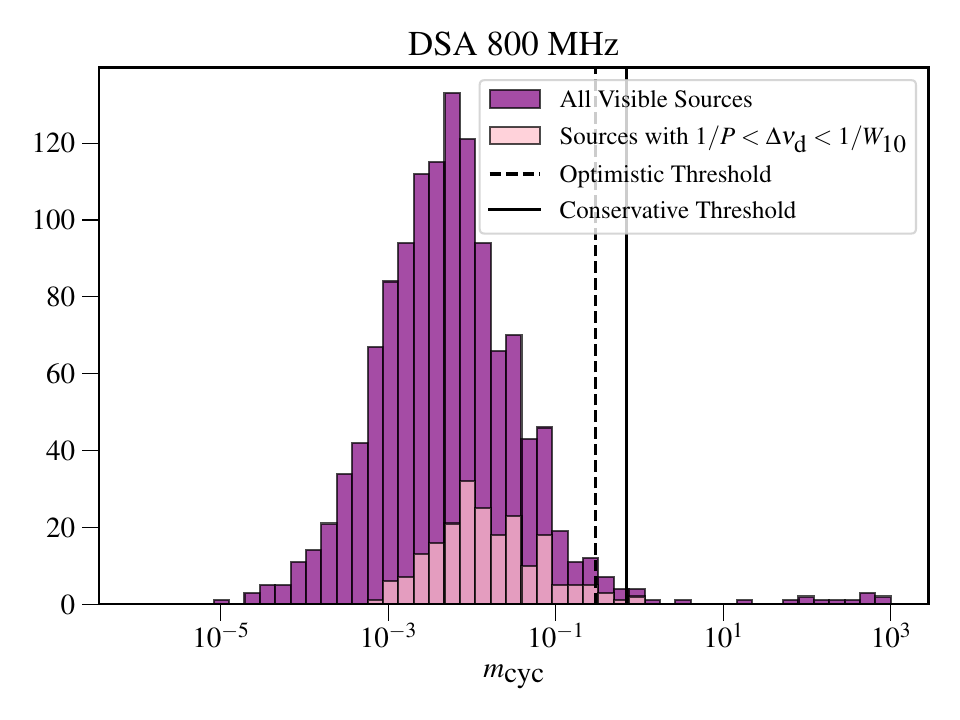}
    \caption{Cyclic merits for all sources analyzed for DSA at 800 MHz (purple), and the subset of sources in the full deconvolution regime (pink). Plot description same as that of Figure 3.}
    \label{fig:dsa_merit}
\end{minipage}
\end{figure}

The capability of the DSA is limited by its frequency coverage, from 0.7-2 GHz. Despite this, it provides a significant increase in sources for which cyclic spectroscopy is viable, specifically in resolving scintles in canonical pulsars. Our survey showed 399 pulsars for which scintles could be resolved. The DSA could prove to be the most useful telescope for cyclic spectroscopy \citep{Turner_2026}. There are three sources that pass the conservative threshold: J1824-1945, J1844+00, and J1845-0743. 13 sources pass the optimistic threshold within the full deconvolution regime. The scintle resolvability is shown in Figure \ref{fig:dsa_scintle} and the histogram for all sources is Figure \ref{fig:dsa_merit}.

\subsection{MeerKAT 600 MHz}
\begin{figure}[h]
\begin{minipage}{.46\textwidth}
    \centering
    \includegraphics[width=0.9\linewidth, left]{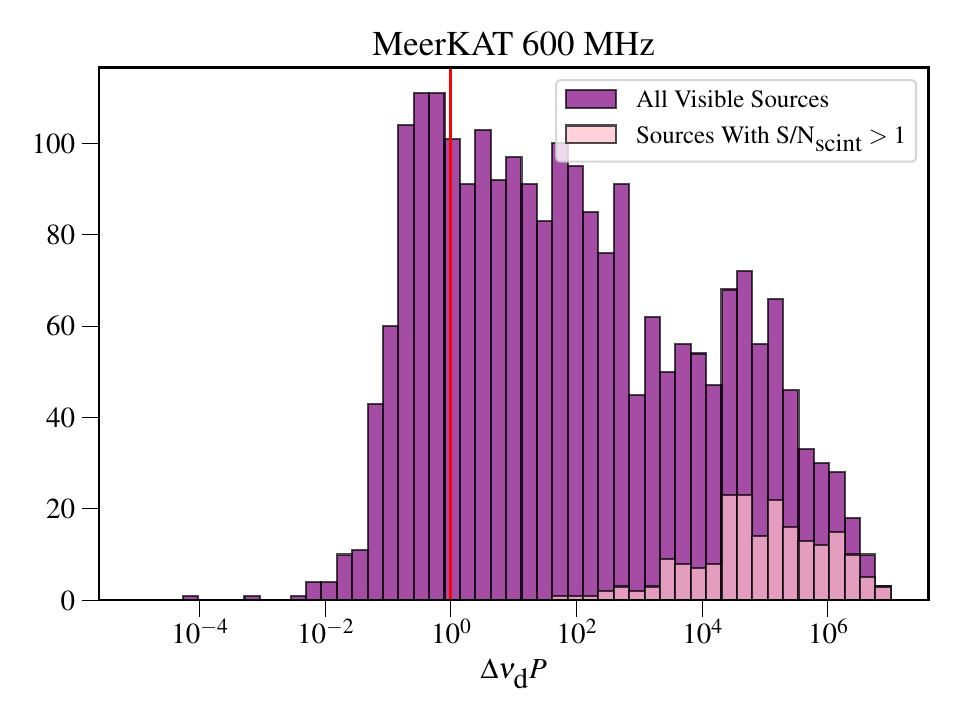}
    \caption{Expected scintle resolvability for all sources analyzed for MeerKAT at 600 MHz (purple), and the subset of sources with S/N$>$1 (pink). Plot description same as that of Figure 2.}
    \label{fig:meerkat_scintle}
\end{minipage}
\hspace{1cm}
\begin{minipage}{.46\textwidth}
    \centering
    \includegraphics[width=0.9\linewidth, right]{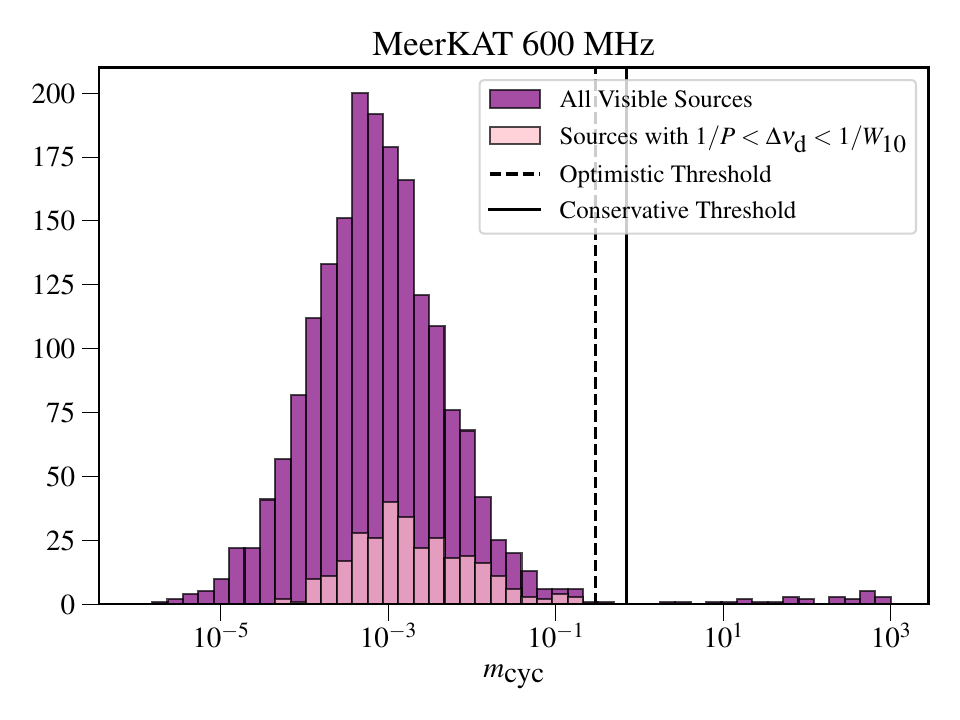}
    \caption{Cyclic merits for all sources analyzed for MeerKAT at 600 MHz (purple), and the subset of sources in the full deconvolution regime (pink). Plot description same as that of Figure 3.}
    \label{fig:meerkat_merit}
\end{minipage}
\end{figure}

MeerKAT has great sensitivity, as well as extensive coverage at lower observing frequencies. Most cyclic deconvolution-based studies done with the Square Kilometre Array (SKA) will have to use the SKA-Low pulsar precursor\textrm{$-$}MWA\textrm{$-$}rather than MeerKAT, which is the SKA-Mid pathfinder. SKA-Mid will be able to observe as low as 350 MHz, which will make it a highly compelling instrument for cyclic spectroscopy. MeerKAT and SKA-Mid will be excellent for scintillation studies using cyclic spectroscopy-processed data. The scintle resolvability is shown in Figure \ref{fig:meerkat_scintle} and the histogram for all sources is shown in Figure \ref{fig:meerkat_merit}. 

\subsection{FAST 550 MHz}
\begin{figure}[h]
\begin{minipage}{.46\textwidth}
    \centering
    \includegraphics[width=0.9\linewidth, left]{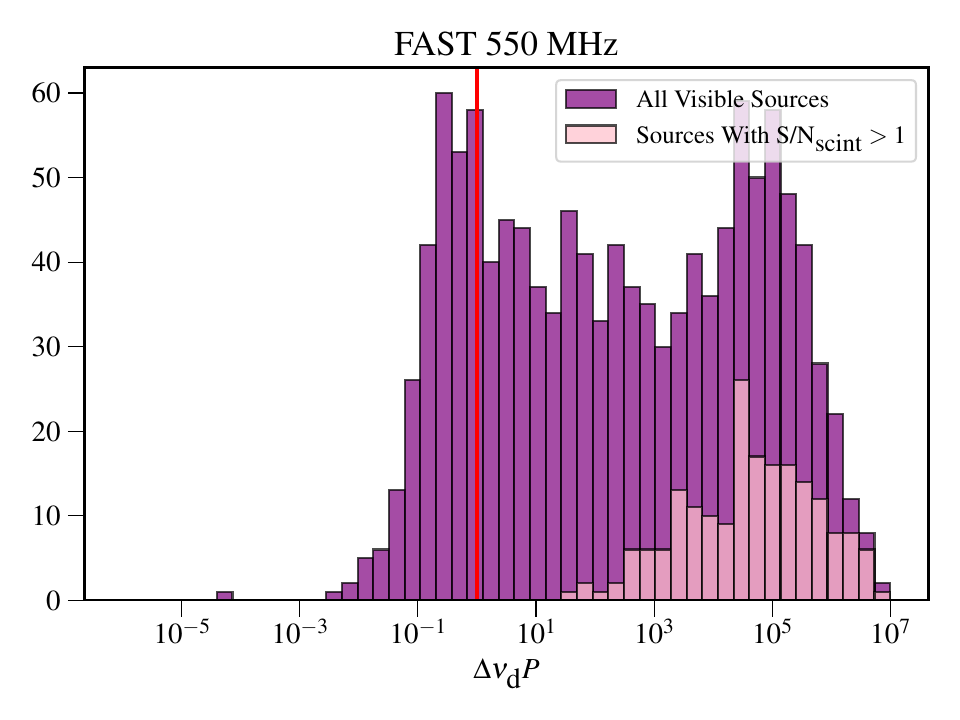}
    \caption{Expected scintle resolvability for all sources analyzed for FAST at 550 MHz (purple), and the subset of sources with S/N$>$1 (pink). Plot description same as that of Figure 2.}
    \label{fig:fast_scintle}
\end{minipage}
\hspace{1cm}
\begin{minipage}{.46\textwidth}
    \centering
    \includegraphics[width=0.9\linewidth, left]{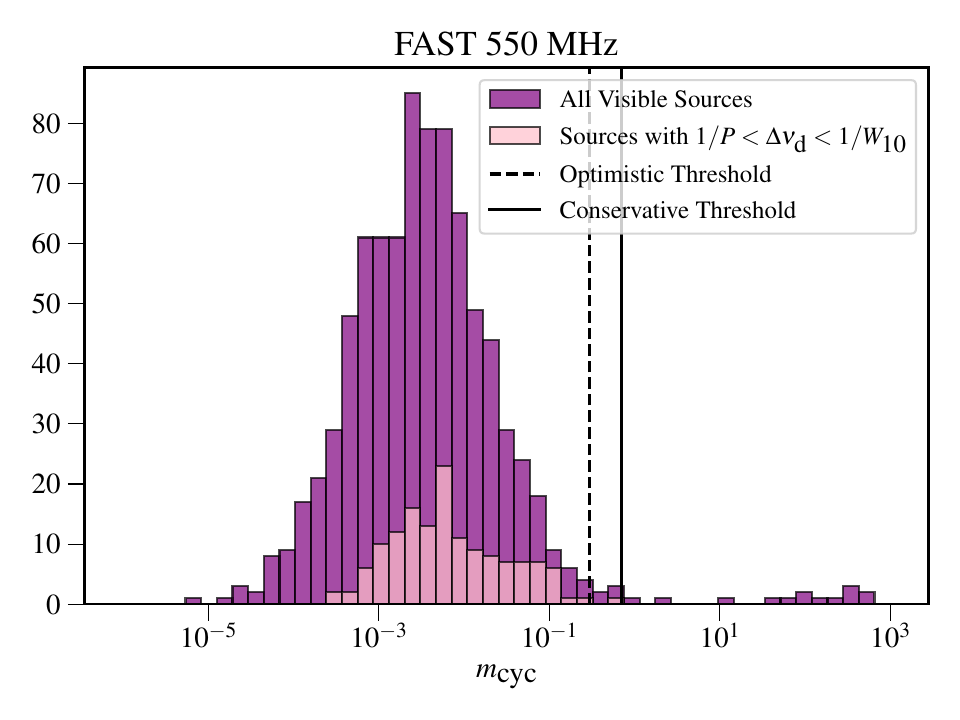}
    \caption{Cyclic merits for all sources analyzed for FAST at 550 MHz (purple), and the subset of sources in the full deconvolution regime (pink). Plot description same as that of Figure 3.}
    \label{fig:fast_merit}
\end{minipage}
\end{figure}

FAST is quite promising for scintle resolvability, with 193 sources having resolvable scintles in our survey. Lower frequency receivers exist, but they may not be available outside of commissioning and early science evaluations. We studied 550 MHz for this telescope because FAST is significantly more sensitive than other telescopes at this frequency. FAST also received an ultra wideband receiver from 0.5-3 GHz that appears very promising for science use cases \citep{Zhang_2023}. FAST has three sources passing the optimistic threshold for the cyclic figure of merit and one passing the conservative threshold (J1844+00), if considering only the full deconvolution regime. The scintle resolvability is shown in Figure \ref{fig:fast_scintle} and the full histogram of all sources analyzed is Figure \ref{fig:fast_merit}.

\subsection{MWA 150 and 200 MHz}
\begin{figure}[h]
\begin{minipage}{.46\textwidth}
    \centering
    \includegraphics[width=0.9\linewidth, right]{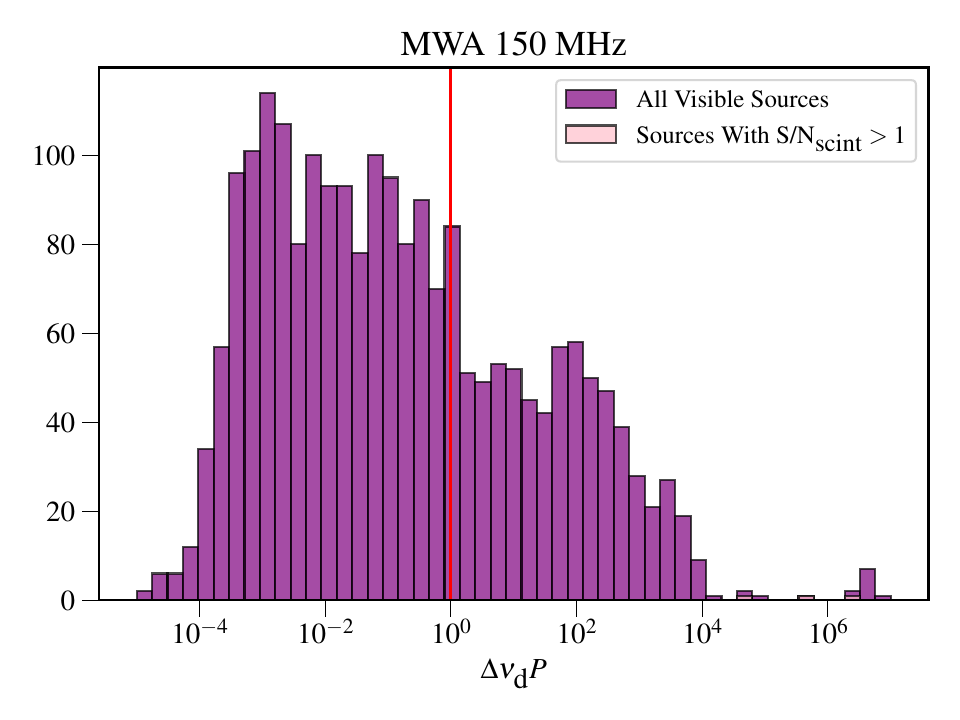}
    \caption{Expected scintle resolvability for all sources analyzed for MWA at 150 MHz (purple), and the subset of sources with S/N$>$1 (pink). Plot description same as that of Figure 2.}
    \label{fig:mwa_150_scintle}
\end{minipage}
\begin{minipage}{.46\textwidth}
    \centering
    \includegraphics[width=0.9\linewidth, right]{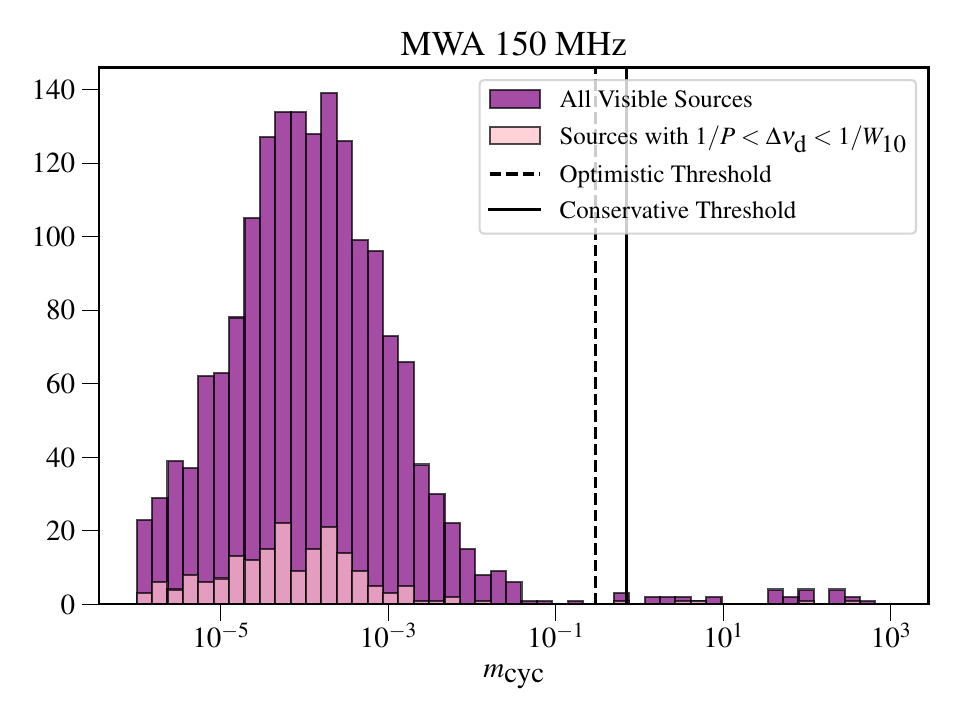}
    \caption{Cyclic merits for all sources analyzed for MWA at 150 MHz (purple), and the subset of sources in the full deconvolution regime (pink). Plot description same as that of Figure 3.}
    \label{fig:mwa_150_merit}
\end{minipage}
\end{figure}

\begin{figure}[h]
\begin{minipage}{.46\textwidth}
    \centering
    \includegraphics[width=0.9\linewidth, left]{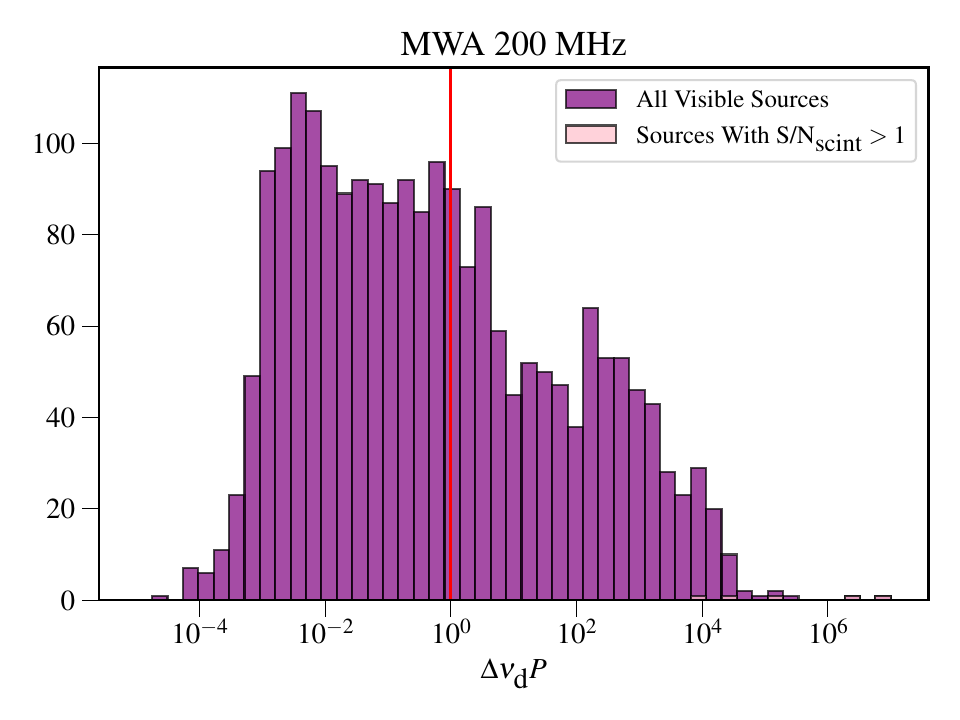}
    \caption{Expected scintle resolvability for all sources analyzed for MWA at 200 MHz (purple), and the subset of sources with S/N$>$1 (pink). Plot description same as that of Figure 2.}
    \label{fig:mwa_200_scintle}
\end{minipage}
\begin{minipage}{.46\textwidth}
    \centering
    \includegraphics[width=0.9\linewidth, left]{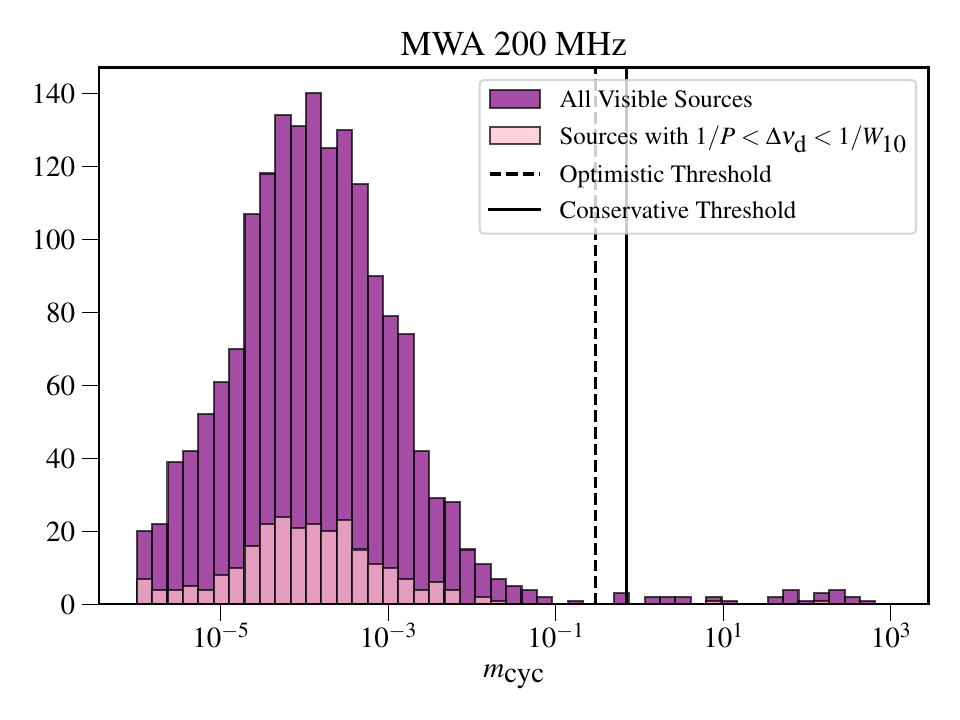}
    \caption{Cyclic merits for all sources analyzed for MWA at 200 MHz (purple), and the subset of sources in the full deconvolution regime (pink). Plot description same as that of Figure 3.}
    \label{fig:mwa_200_merit}
\end{minipage}
\end{figure}

MWA has 6 sources that pass the conservative threshold at 150 MHz (B0523+11, B0540+23, B0626+24, B0833-45, B1857-26, and B1907+23) and 3 at 200 MHz (B053+11, B0540+23, and B0833-45). B0833-45, commonly known as the Vela pulsar, is particularly promising on this instrument. There are 3 sources with resolvable scintles for 150 MHz and 5 for 200 MHz. This could indicate that the best telescopes for cyclic spectroscopy on canonical pulsars, as measured by the cyclic figure of merit, are the worst for resolving scintles. It is more likely that this result is a shortcoming of the cyclic figure of merit, discussed more in the next section. More sensitive telescopes do not show this trend as clearly, and it is not present with MSPs. Cyclic figure of merit relies specifically on S/N and scattering delay, so when these quantities are large, the cyclic figure of merit will be large, even when the scattering delay is so large that the pulsar cannot be resolved. Lower frequency receivers could mitigate these issues with the higher flux densities of pulsars that they tend to detect, hence why we advocate for more of these receivers for upcoming telescopes. The scintle resolvability for the frequencies are shown in Figures \ref{fig:mwa_150_scintle} and \ref{fig:mwa_200_scintle}, respectively, and the full histograms are Figures \ref{fig:mwa_150_merit} and \ref{fig:mwa_200_merit}.

\subsection{LWA 74 MHz}
\begin{figure}[h]
\begin{minipage}{.46\textwidth}
    \centering
    \includegraphics[width=0.9\linewidth, right]{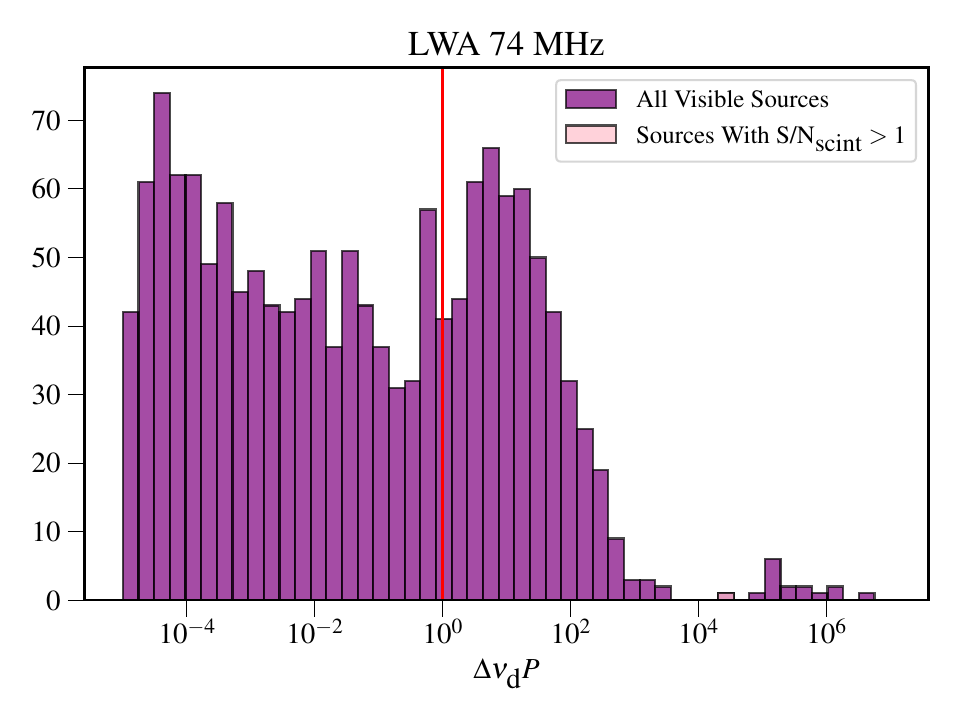}
    \caption{Expected scintle resolvability for all sources analyzed for LWA at 74 MHz (purple), and the subset of sources with S/N$>$1 (pink). Plot description same as that of Figure 2.}
    \label{fig:lwa_scintle}
\end{minipage}
\hspace{1cm}
\begin{minipage}{.46\textwidth}
    \centering
    \includegraphics[width=0.9\linewidth, right]{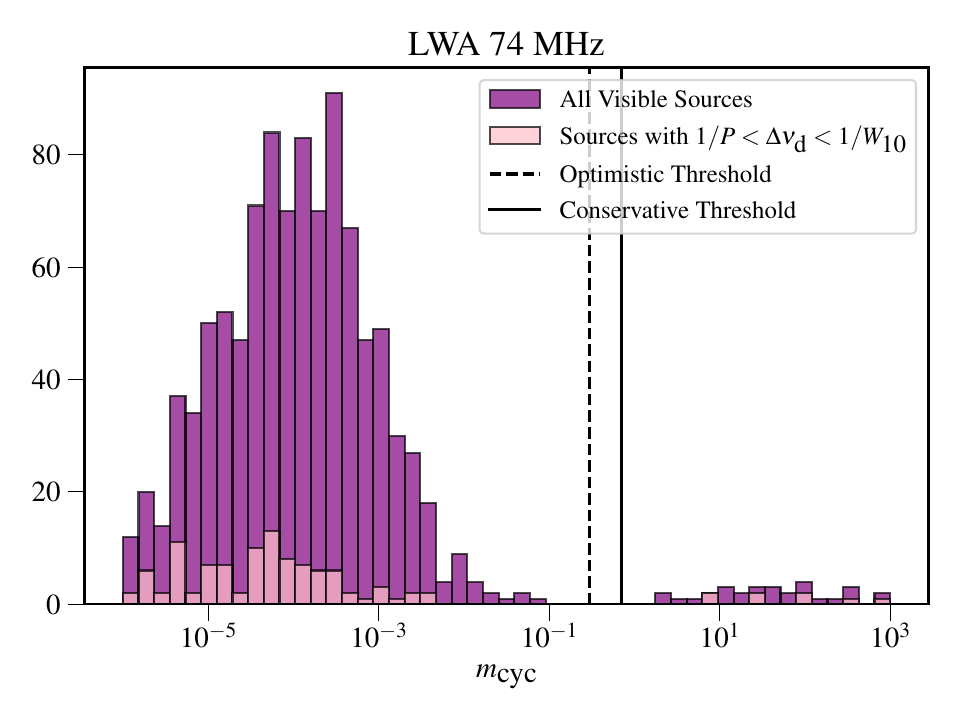}
    \caption{Cyclic merits for all sources analyzed for LWA at 74 MHz (purple), and the subset of sources in the full deconvolution regime (pink). Plot description same as that of Figure 3.}
    \label{fig:lwa_merit}
\end{minipage}
\end{figure}

LWA has one of the lowest observing frequencies of all the telescopes we consider. 11 sources pass the conservative threshold for the cyclic figure of merit in the full deconvolution regime. LWA has only one source for which scintles are resolvable. We believe this shows a discrepancy in the cyclic figure of merit's accuracy. As mentioned, we see a larger number of sources passing the conservative threshold in the full deconvolution regime than of resolvable scintles. We posit that this is a shortcoming of the cyclic figure of merit: it overreports the number of sources that could be deconvolved at low frequencies. The scintle resolvability is shown in Figure \ref{fig:lwa_scintle} and the histogram of all sources analyzed is Figure \ref{fig:lwa_merit}.

\subsection{LOFAR 150 and 200 MHz}
\begin{figure}[h]
\begin{minipage}{.46\textwidth}
    \centering
    \includegraphics[width=0.9\linewidth, left]{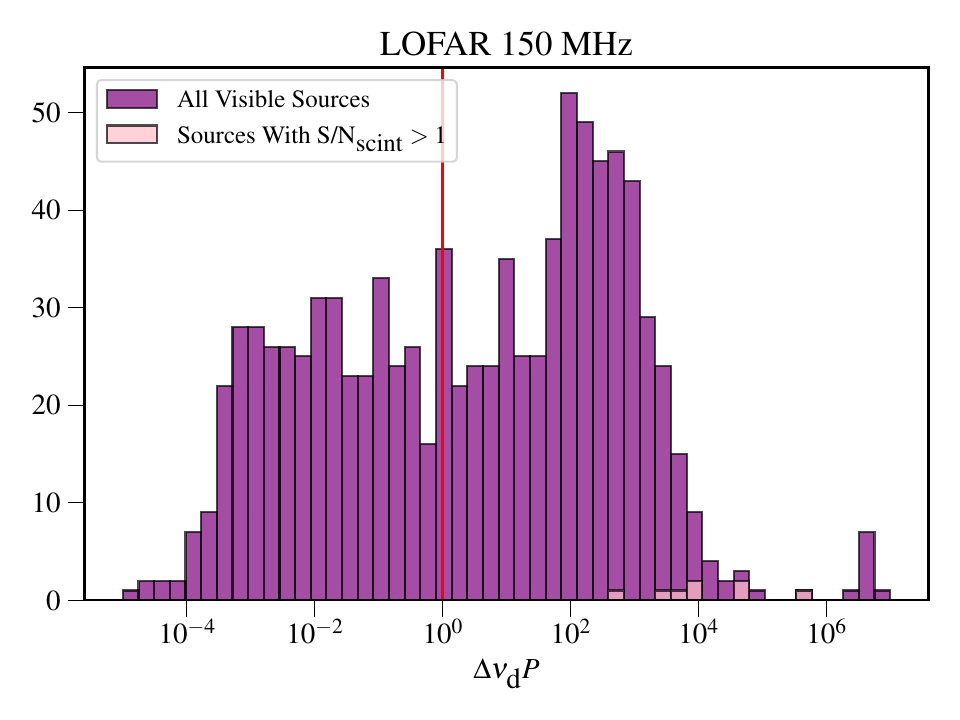}
    \caption{Expected scintle resolvability for all sources analyzed for LOFAR at 150 MHz (purple), and the subset of sources with S/N$>$1 (pink). Plot description same as that of Figure 2.}
    \label{fig:lofar_150_scintle}
\end{minipage}
\hspace{1cm}
\begin{minipage}{.46\textwidth}
    \centering
    \includegraphics[width=0.9\linewidth, left]{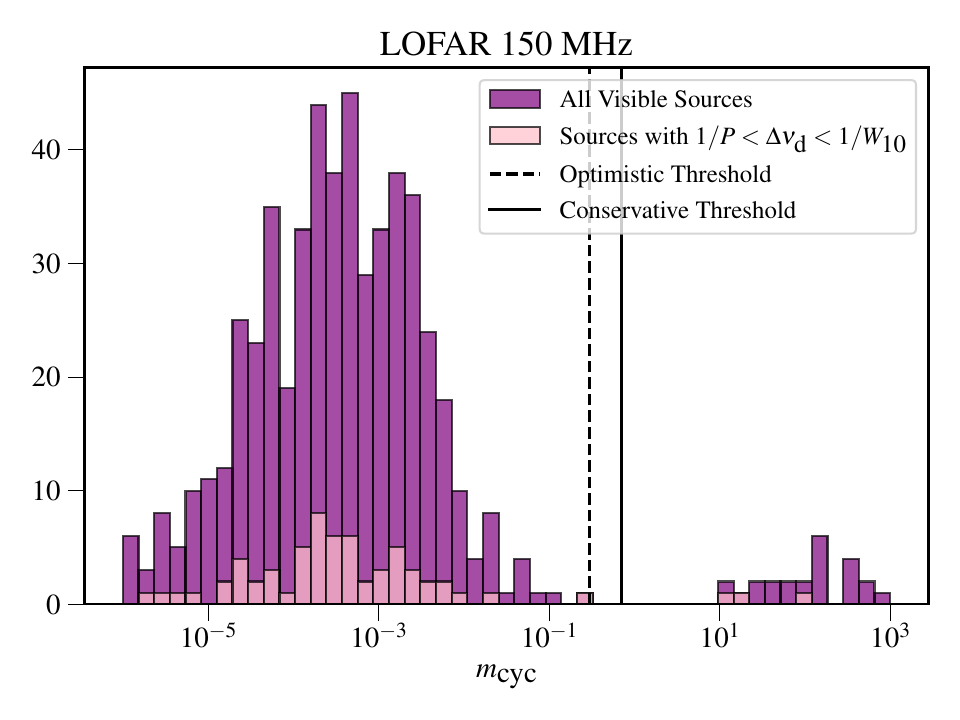}
    \caption{Cyclic merits for all sources analyzed for LOFAR at 150 MHz (purple), and the subset of sources in the full deconvolution regime (pink). Plot description same as that of Figure 3.}
    \label{fig:lofar_150_merit}
\end{minipage}
\end{figure}

\begin{figure}[h]
\begin{minipage}{.46\textwidth}
    \centering
    \includegraphics[width=0.9\linewidth, right]{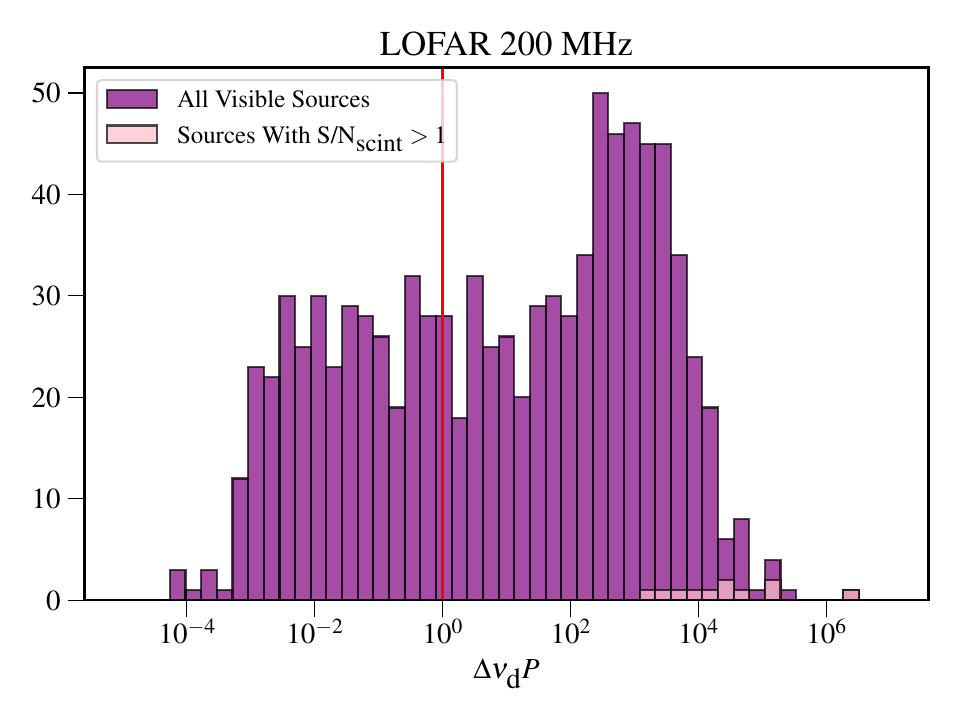}
    \caption{Expected scintle resolvability for all sources analyzed for LOFAR at 200 MHz (purple), and the subset of sources with S/N$>$1 (pink). Plot description same as that of Figure 2.}
    \label{fig:lofar_200_scintle}
\end{minipage}
\hspace{1cm}
\begin{minipage}{.46\textwidth}
    \centering
    \includegraphics[width=0.9\linewidth, right]{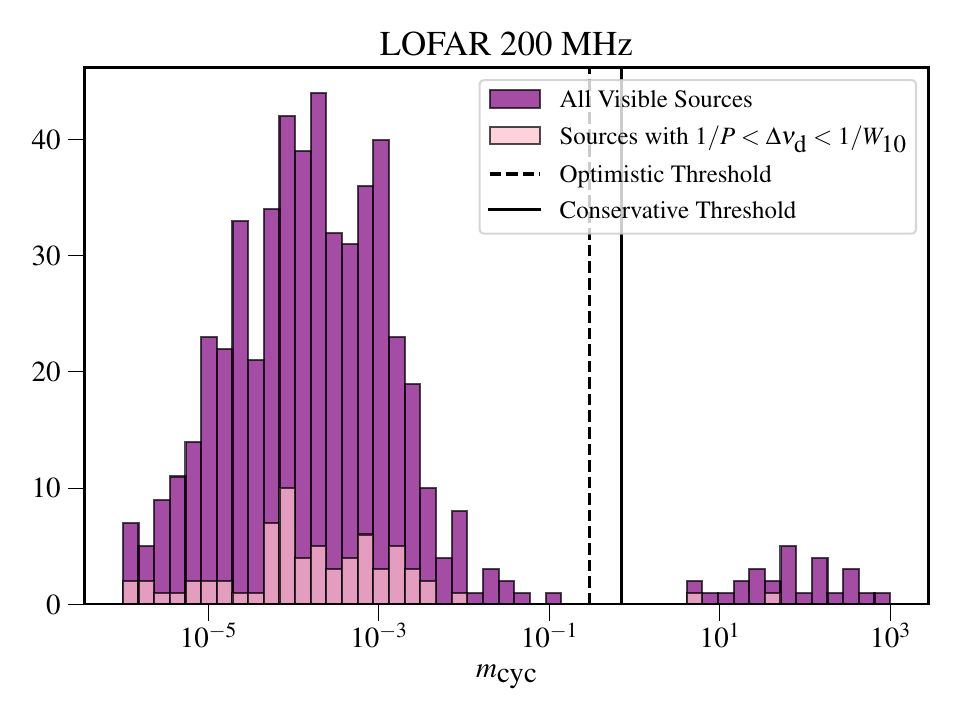}
    \caption{Cyclic merits for all sources analyzed for LOFAR at 200 MHz (purple), and the subset of sources in the full deconvolution regime (pink). Plot description same as that of Figure 3.}
    \label{fig:lofar_200_merit}
\end{minipage}
\end{figure}

While LOFAR is one of the best instruments to use for cyclic spectroscopy on millisecond pulsars, its results for canonicals are not any more impressive than the other combinations we studied\textrm{$-$}2 sources passing the conservative threshold for 200 MHz (B0523+11 and B0540+23) and 4 for 150 MHz (B0523+11, B0540+23, B0626+24, and B1857-26). The scintle resolveabilities are shown in Figures \ref{fig:lofar_150_scintle} and \ref{fig:lofar_200_scintle}, and the full histograms for all sources are shown in Figures \ref{fig:lofar_150_merit} and \ref{fig:lofar_200_merit}. 

\subsection{NenuFAR 48 and 85 MHz}
\begin{figure}[h]
\begin{minipage}{.46\textwidth}
    \centering
    \includegraphics[width=0.9\linewidth, left]{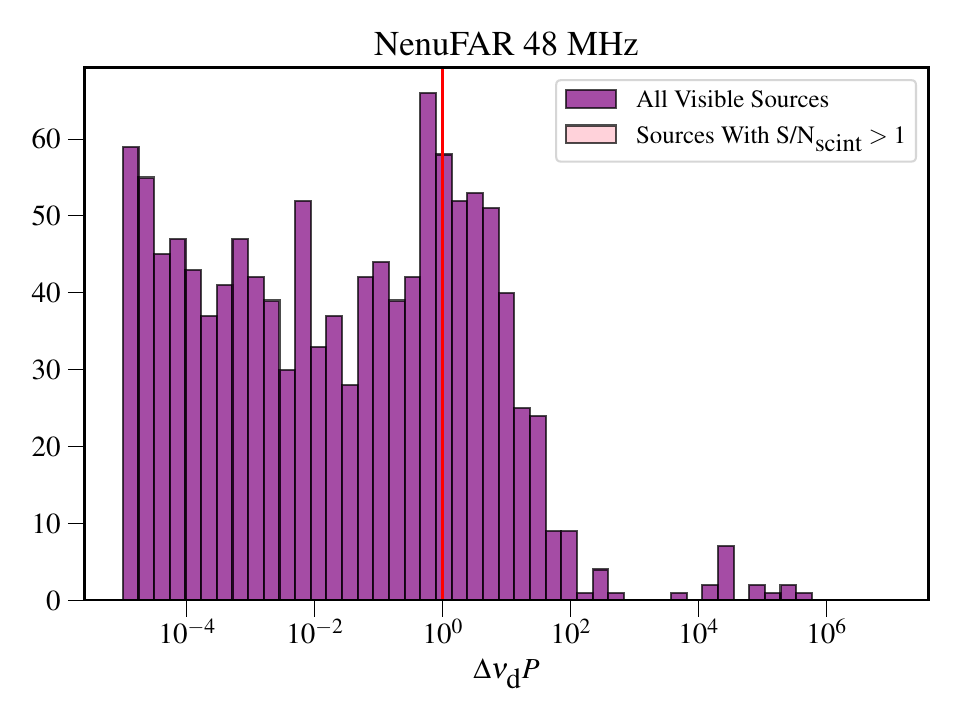}
    \caption{Expected scintle resolvability for all sources analyzed for NenuFAR at 48 MHz (purple), and the subset of sources with S/N$>$1 (pink). Plot description same as that of Figure 2.}
    \label{fig:nenufar_48_scintle}
\end{minipage}
\hspace{1cm}
\begin{minipage}{.46\textwidth}
    \centering
    \includegraphics[width=0.9\linewidth, left]{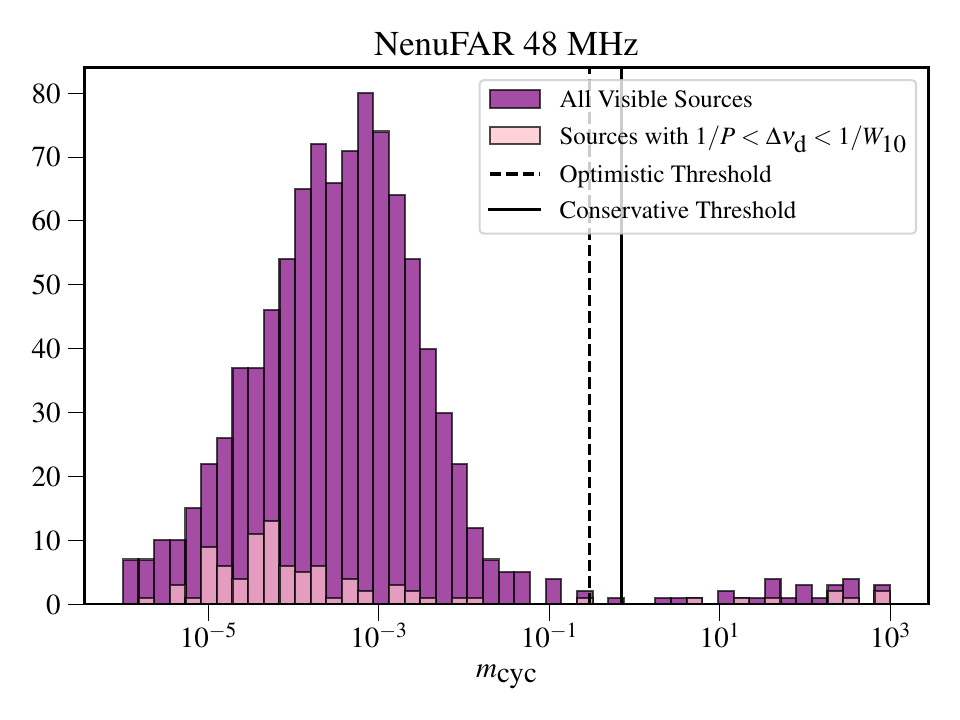}
    \caption{Cyclic merits for all sources analyzed for NenuFAR at 48 MHz (purple), and the subset of sources in the full deconvolution regime (pink). Plot description same as that of Figure 3.}
    \label{fig:nenufar_48_merit}
\end{minipage}
\end{figure}

\begin{figure}[h]
\begin{minipage}{.46\textwidth}
    \centering
    \includegraphics[width=0.9\linewidth, right]{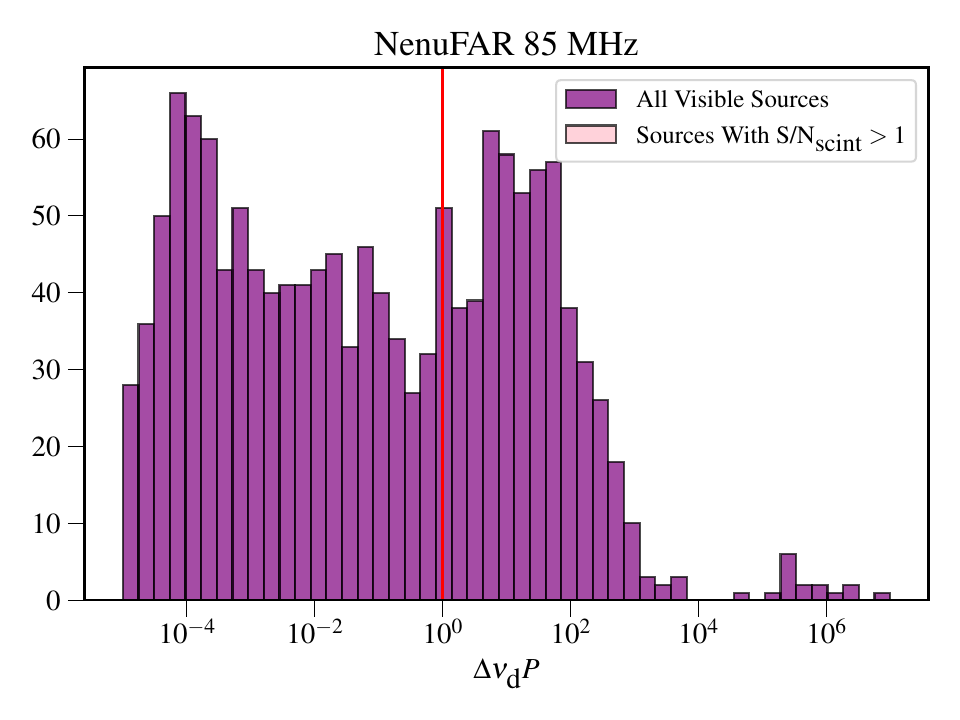}
    \caption{Expected scintle resolvability for all sources analyzed for NenuFAR at 85 MHz (purple), and the subset of sources with S/N$>$1 (pink). Plot description same as that of Figure 2.}
    \label{fig:nenufar_85_scintle}
\end{minipage}
\hspace{1cm}
\begin{minipage}{.46\textwidth}
    \centering
    \includegraphics[width=0.9\linewidth, right]{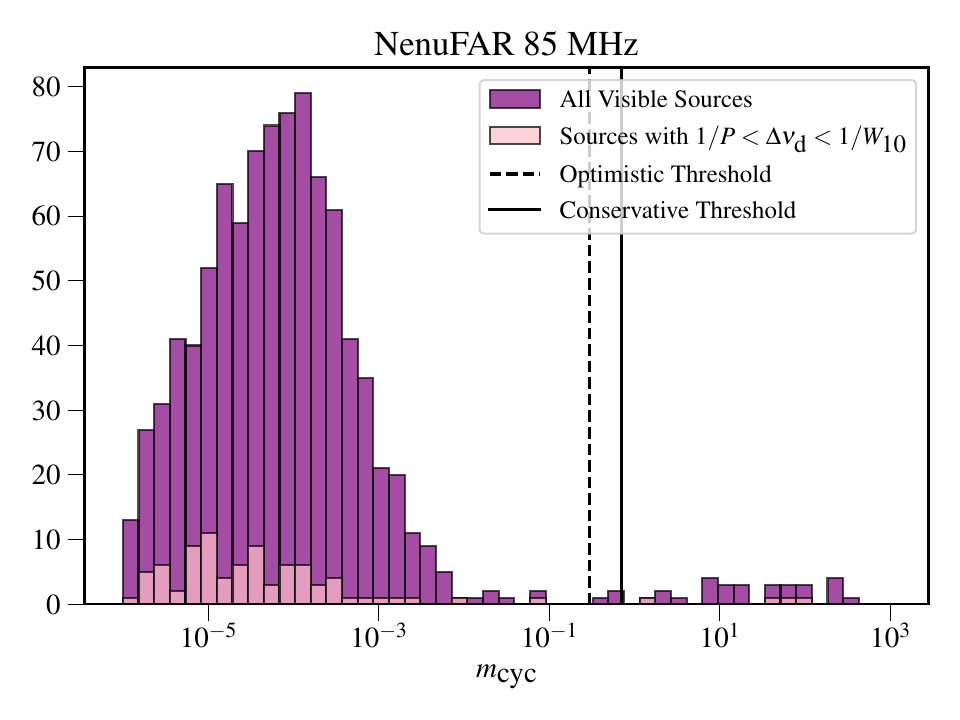}
    \caption{Cyclic merits for all sources analyzed for NenuFAR at 85 MHz (purple), and the subset of sources in the full deconvolution regime (pink). Plot description same as that of Figure 3.}
    \label{fig:nenufar_85_merit}
\end{minipage}
\end{figure}

NenuFAR is incredibly sensitive and has the lowest observing frequency of all the telescopes we considered, though its results are not much better than other telescopes. Interestingly, the 48 MHz band has two more sources that pass the conservative threshold than the 85 MHz band, suggesting the prediction expressed in \cite{Turner_2026} that lower frequencies make cyclic deconvolution very difficult may not be an issue. The small number of sources indicates we would need more data to confirm this. We were able to address the issue of using a simple power law to calculate flux density, and thus overestimating our ability to use cyclic spectroscopy on these sources with NenuFAR. We used whichever law was recommended by the \textsc{pulsar-spectra} package, unless there was no data for these pulsars in the \textsc{pulsar-spectra} database, in which case we opted to use a power law with a $-$1.7 spectral index. Thus, some pulsars’ prospects for cyclic spectroscopy may still be overestimated, but we have minimized this issue as much as possible. It is also worth noting that the S/N per scintle of all the sources within the full deconvolution regime is on the order of 10$^{-8}$, calling into question the actual viability of this technique on this instrument. NenuFAR is a lower-frequency telescope, so it is possible that there are some sources at low frequencies that have large cyclic merits, but are not actually detectable due to low S/N. The scintle resolvability is shown in Figures \ref{fig:nenufar_48_scintle} and \ref{fig:nenufar_85_scintle}, and the histograms of all sources analyzed are Figures \ref{fig:nenufar_48_merit} and \ref{fig:nenufar_85_merit}.

\subsection{CHIME 450 MHz}
\begin{figure}[h]
\begin{minipage}{.46\textwidth}
    \centering
    \includegraphics[width=0.9\linewidth, left]{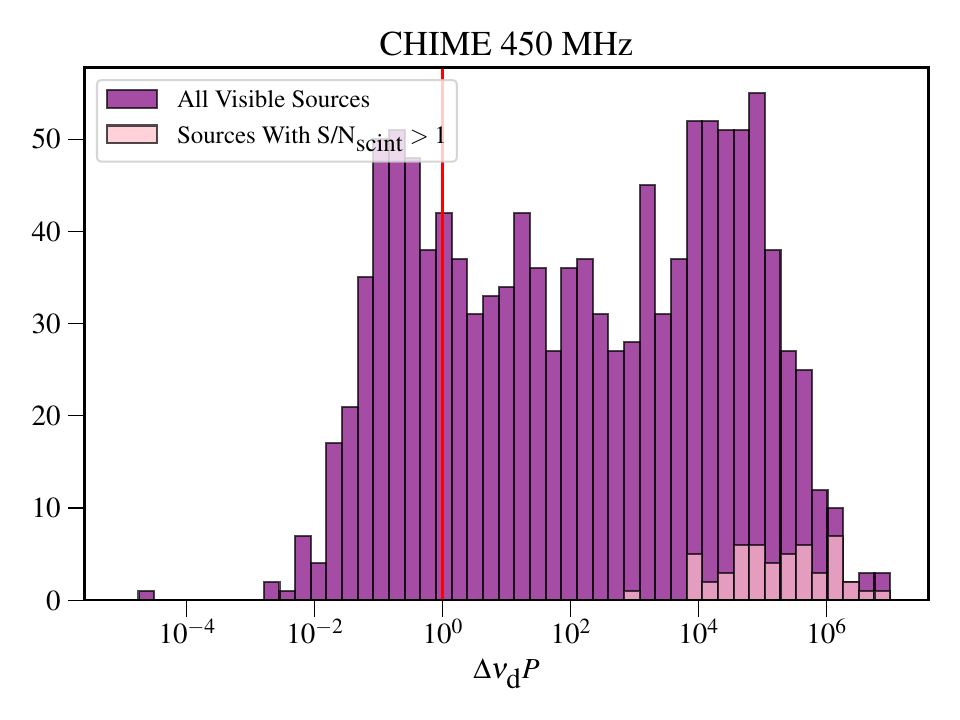}
    \caption{Expected scintle resolvability for all sources analyzed for CHIME at 450 MHz (purple), and the subset of sources with S/N$>$1 (pink). Plot description same as that of Figure 2.}
    \label{fig:chime_scintle}
\end{minipage}
\hspace{1cm}
\begin{minipage}{.46\textwidth}
    \centering
    \includegraphics[width=0.9\linewidth, left]{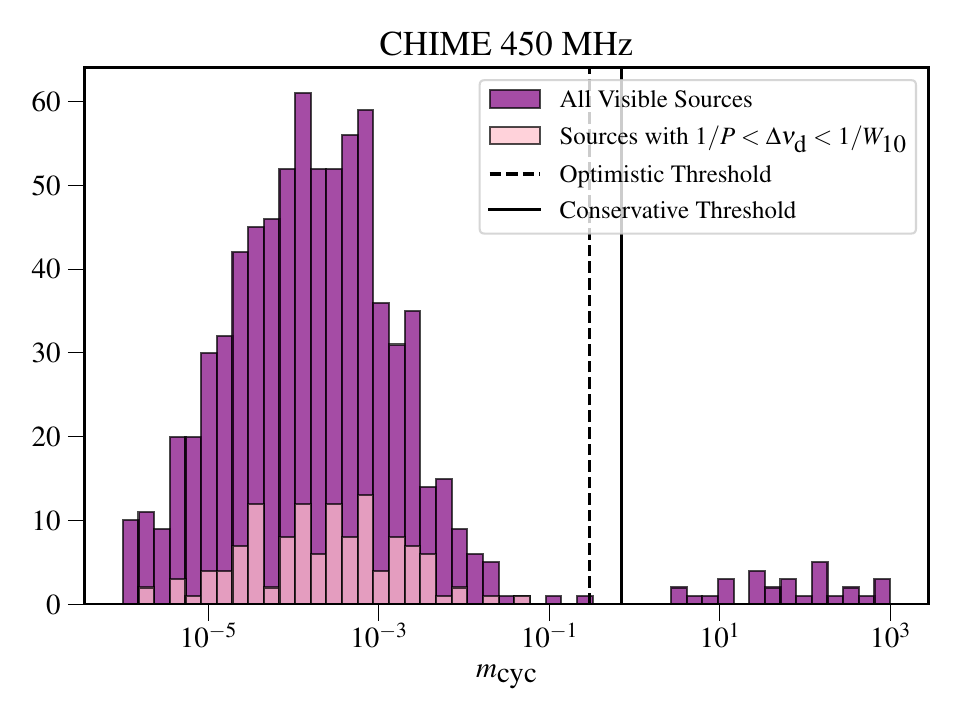}
    \caption{Cyclic merits for all sources analyzed for CHIME at 450 MHz (purple), and the subset of sources in the full deconvolution regime (pink). Plot description same as that of Figure 3.}
    \label{fig:chime_merit}
\end{minipage}
\end{figure}

CHIME observes in a relatively sparse frequency range for cyclic deconvolution. It also has relatively low sensitivity and thus may not be able to detect some sources that are within their full deconvolution regime. CHIME also has a short observing time of only 10 minutes per day, further limiting the instrument to recovering pulse broadening functions rather than studying full dynamic wavefields. CHIME does not have any sources passing the conservative threshold for the cyclic figure of merit. 53 sources have resolvable scintles when observed with CHIME. The scintle resolvability is shown in Figure \ref{fig:chime_scintle} and the histogram of all sources analyzed is Figure \ref{fig:chime_merit}.

\subsection{uGMRT 150, 200, and 350 MHz}
\begin{figure}[h]
\begin{minipage}{.46\textwidth}
    \centering
    \includegraphics[width=0.9\linewidth, right]{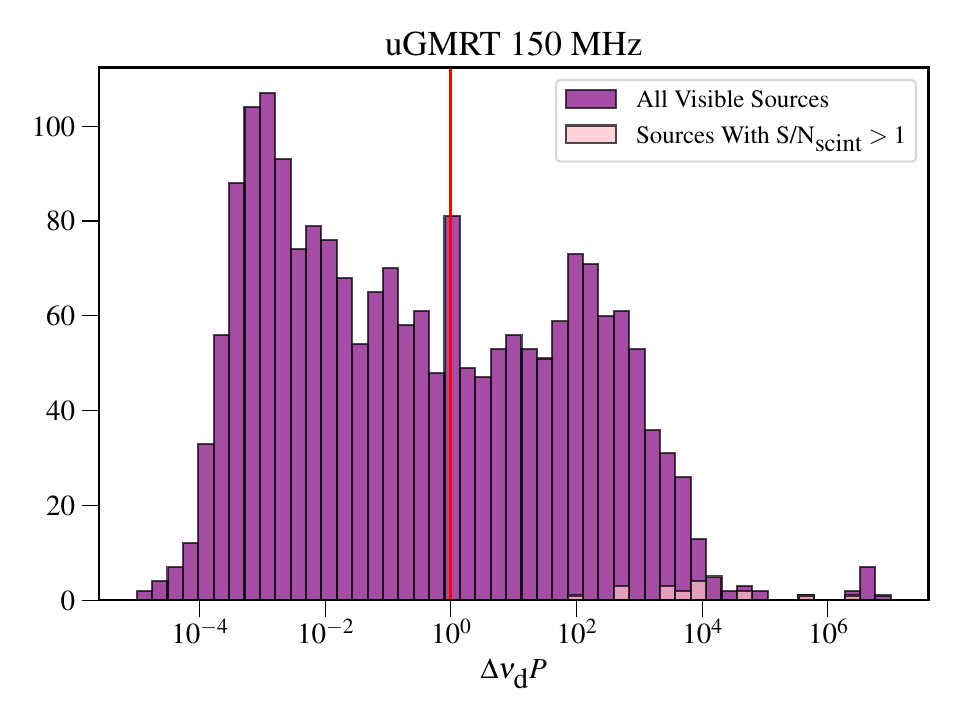}
    \caption{Expected scintle resolvability for all sources analyzed for uGMRT at 150 MHz (purple), and the subset of sources with S/N$>$1 (pink). Plot description same as that of Figure 2.}
    \label{fig:ugmrt_150_scintle}
\end{minipage}
\hspace{1cm}
\begin{minipage}{.46\textwidth}
    \centering
    \includegraphics[width=0.9\linewidth, right]{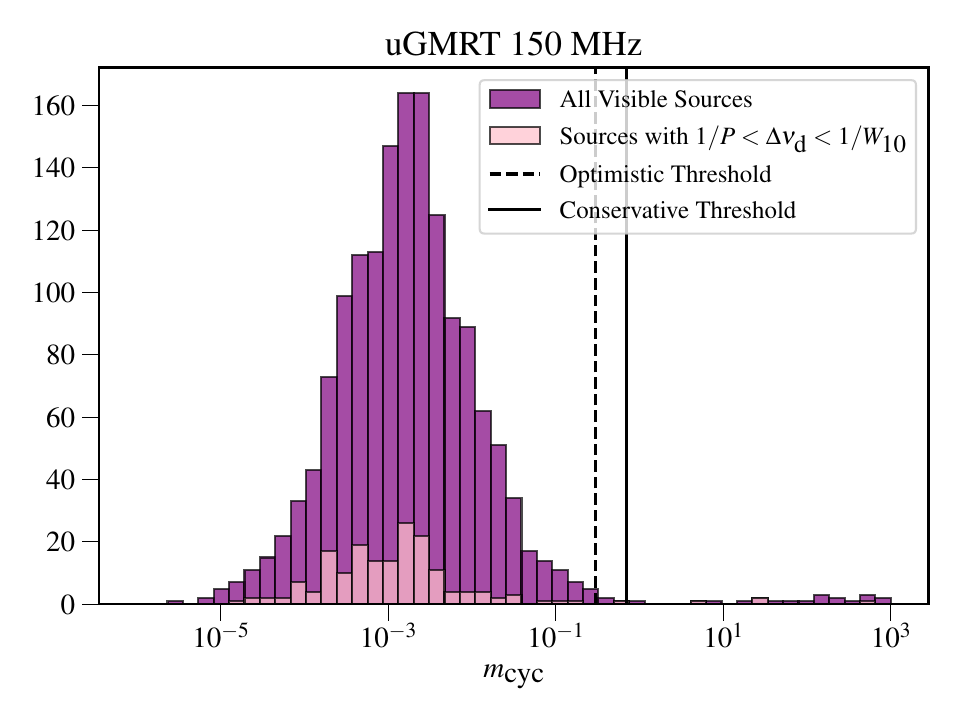}
    \caption{Cyclic merits for all sources analyzed for uGMRT at 150 MHz (purple), and the subset of sources in the full deconvolution regime (pink). Plot description same as that of Figure 3.}
    \label{fig:ugmrt_150_merit}
\end{minipage}
\end{figure}

\begin{figure}[h]
\begin{minipage}{.46\textwidth}
    \centering
    \includegraphics[width=0.9\linewidth, left]{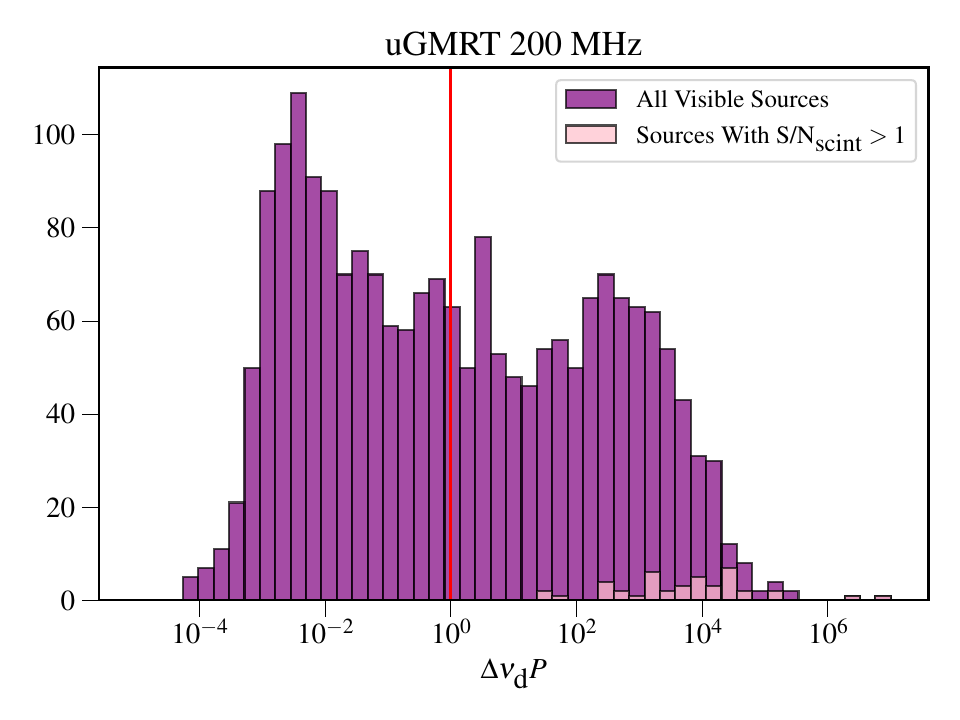}
    \caption{Expected scintle resolvability for all sources analyzed for uGMRT at 200 MHz (purple), and the subset of sources with S/N$>$1 (pink). Plot description same as that of Figure 2.}
    \label{fig:ugmrt_200_scintle}
\end{minipage}
\hspace{1cm}
\begin{minipage}{.46\textwidth}
    \centering
    \includegraphics[width=0.9\linewidth, left]{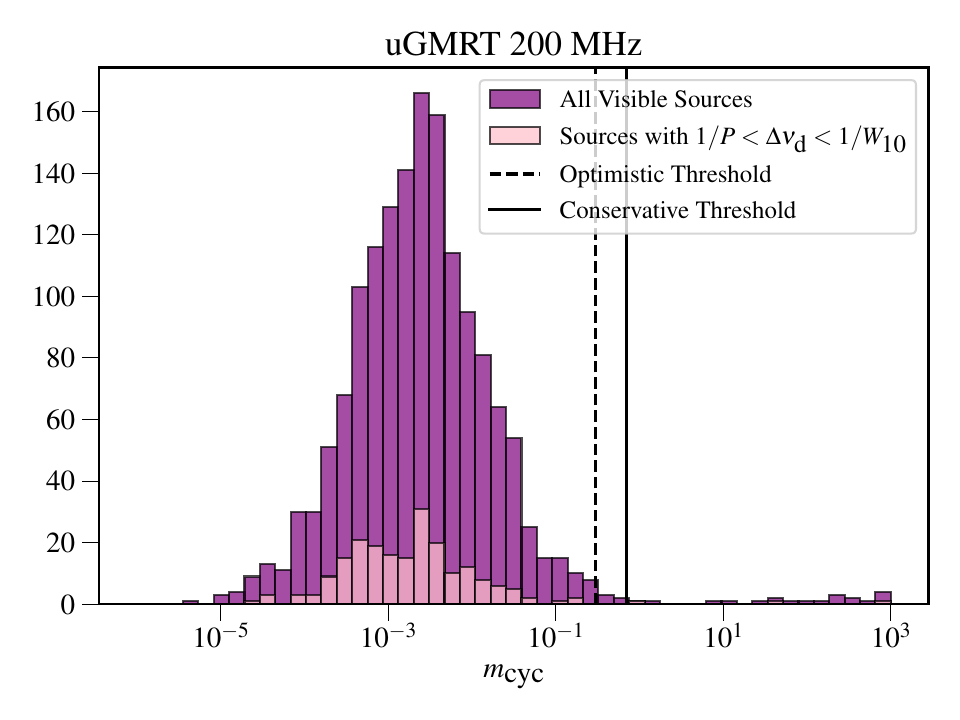}
    \caption{Cyclic merits for all sources analyzed for uGMRT at 200 MHz (purple), and the subset of sources in the full deconvolution regime (pink). Plot description same as that of Figure 3.}
    \label{fig:ugmrt_200_merit}
\end{minipage}
\end{figure}

\begin{figure}[h]
\begin{minipage}{.46\textwidth}
    \centering
    \includegraphics[width=0.9\linewidth, right]{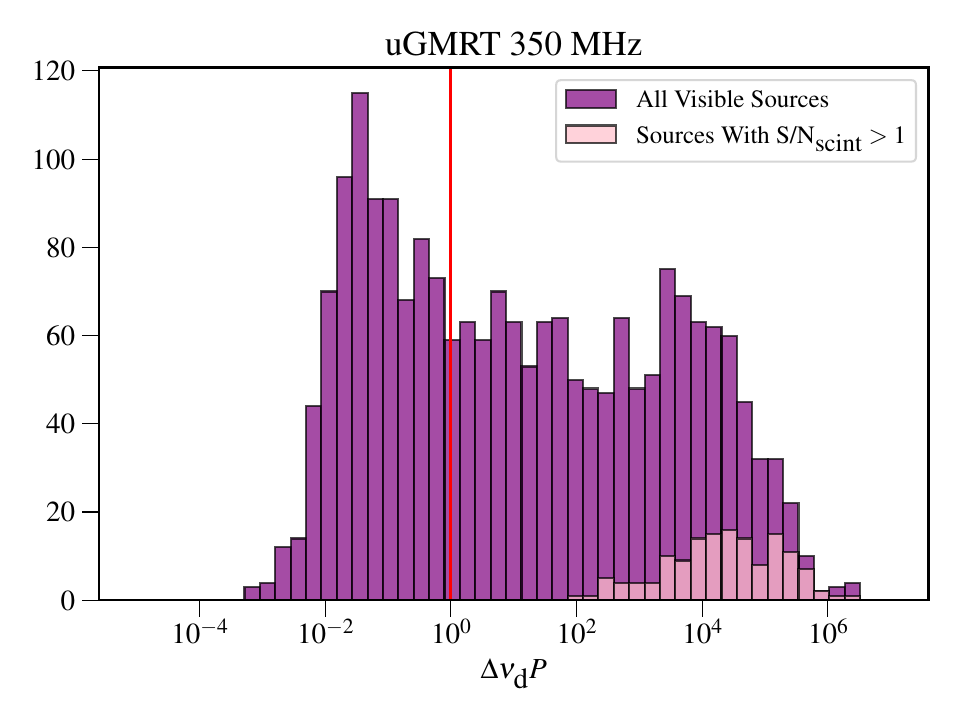}
    \caption{Expected scintle resolvability for all sources analyzed for uGMRT at 350 MHz (purple), and the subset of sources with S/N$>$1 (pink). Plot description same as that of Figure 2.}
    \label{fig:ugmrt_350_scintle}
\end{minipage}
\hspace{1cm}
\begin{minipage}{.46\textwidth}
    \centering
    \includegraphics[width=0.9\linewidth, right]{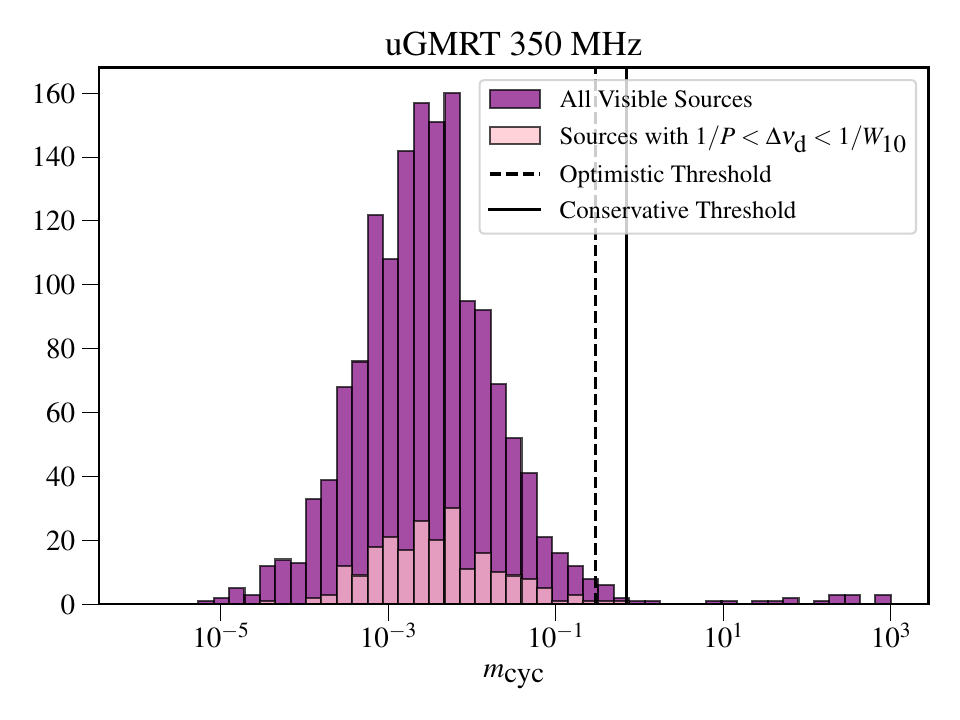}
    \caption{Cyclic merits for all sources analyzed for uGMRT at 350 MHz (purple), and the subset of sources in the full deconvolution regime (pink). Plot description same as that of Figure 3.}
    \label{fig:ugmrt_350_merit}
\end{minipage}
\end{figure}

uGMRT is a useful instrument for resolving scintles with cyclic spectroscopy, especially considering the instrument was not created for this purpose. There are 6 sources passing the conservative threshold at 150 MHz, 3 at 200 MHz, and 1 at 350 MHz. uGMRT is also promising for resolving scintles, with 17 sources at 150 MHz, 42 at 200 MHz, and 144 at 350 MHz. We may be able to take advantage of uGMRT’s subarraying capabilities to perform simultaneous scans across all three observing bands, which would save time and survey more frequencies than using only one receiver. However, three simultaneous observations would reduce the sensitivity by a large factor, a trade-off which we may not be able to make. The scintle resolveabilities for 150, 200, and 350 MHz are shown in Figures \ref{fig:ugmrt_150_scintle}, \ref{fig:ugmrt_200_scintle}, and \ref{fig:ugmrt_350_scintle} respectively, and histograms of all sources analyzed are Figures \ref{fig:ugmrt_150_merit}, \ref{fig:ugmrt_200_merit}, and \ref{fig:ugmrt_350_merit}.

\section{Conclusions}
\label{sec:conclusions}
We have calculated cyclic merits for 2609 pulsars across 15 telescope-frequency combinations to determine whether cyclic spectroscopy is a viable technique for canonical pulsars. We found that the LWA and NenuFAR telescopes, at 74 and 48 MHz, were the highest-yielding receivers for cyclic spectroscopy with 11 and 12 sources within the full deconvolution regime passing the conservative threshold respectively. Due to the wider profiles of canonical pulsars as compared to MSPs, this result is expected; cyclic spectroscopy works best on pulsars whose pulse profiles are narrow, so even our telescopes that perform with the most promise may not have equivalent results to those on MSPs. Cyclic merit may be biased to make some pulsars appear like they can have cyclic deconvolution performed when this is not actually true, again making our most high-performing telescopes have potentially unreliable results. The overreporting issue will be explored more in future work, but for now we advise that these results are tenuous, and a recalibration of the cyclic figure of merit may be in order, at least for use with the canonical population. More promising in the context of canonical pulsars is the prospect of using cyclic spectroscopy backends to resolve scintles. The best telescopes for this were MeerKAT (600 MHz) and DSA (800 MHz), implying that higher frequency observations are best for resolving scintles. Pulsar science would benefit from more telescopes developing real-time cyclic spectroscopy backends, as the GBT currently has. Future work will include attempting cyclic deconvolution and scintle resolution with real data, both on the canonical population and on MSPs, specifically focusing on use of the telescopes that have the most sources resolving scintles and passing the conservative threshold for the cyclic figure of merit.

\startlongtable
\begin{deluxetable*}{CCCCCCCCCCC}
\setlength{\tabcolsep}{0pt}
\tablecaption{Cyclic Merits for GBT, DSA, MeerKAT, FAST, MWA, \& LWA \label{cyclic_merits}}
\tablehead{\colhead{Pulsar} & \colhead{\text{GBT}$_{350}$} & \colhead{\text{DSA}$_{800}$} & \colhead{\text{MeerKAT}$_{600}$} & \colhead{\text{FAST}$_{550}$} & \colhead{\text{MWA}$_{150}$} & \colhead{\text{MWA}$_{200}$} & \colhead{\text{LWA}$_{74}$} &\colhead{Max Freq} & \colhead{Min Freq}}

\startdata
\textrm{B}0031\textrm{$--$}07 &    393 (212) &   2038 (1098) &     285 (154) &    1546 (833) &    \textrm{---} &    \textrm{---} &      156 (83.9) &        5 (2) &       3 (1) \\
\textrm{B}0138{+}59 &    625 (564) &   3133 (2830) &     \textrm{---} &   2380 (2149) &   \textrm{---} &   \textrm{---} &     65.5 (59.2) &      60 (30) &     34 (17) \\
    \textrm{B}0329{+}54 &  6682 (3385) & 32910 (16672) &   \textrm{---} & 25004 (12667) &   \textrm{---} &   \textrm{---} &     3095 (1568) &       10 (5) &       5 (2) \\
    \textrm{B}0450{+}55 &  1186 (1074) &   6057 (5484) &     \textrm{---} &   4599 (4163) &    \textrm{---} &     \textrm{---} &       327 (296) &     81 (41) &     46 (23) \\
    \textrm{B}0450\textrm{$--$}18 &  2015 (2590) & 13137 (16887) &   1875 (2410) &  \textrm{---} &     356 (458) &     418 (537) &       414 (532) &      76 (38) &     41 (21) \\
    \textrm{B}0523{+}11 &  18.7 (9.35) &    106 (52.9) &   14.9 (7.45) &   80.0 (40.0) &   6.07 (3.04) &    6.79 (3.4) & \textrm{---}  &   273 (137) &    146 (73) \\
    \textrm{B}0525{+}21 &  9.89 (4.97) &   58.7 (29.5) &    8.3 (4.17) &   44.4 (22.3) &         <0.01 &         <0.01 &           <0.01 &     143 (72) &     74 (37) \\
    \textrm{B}0540{+}23 &    346 (174) &   2109 (1061) &     299 (150) &    1594 (802) &    112 (56.2) &    129 (64.7) &      134 (67.5) &    221 (110) &    120 (60) \\
\enddata
\tablecomments{Cyclic merits and upper and lower observing frequency limits on the full deconvolution regime for GBT, DSA, MeerKAT, FAST, MWA, LWA, \& SKA-Low. Some values that may pass either cyclic merit threshold, optimistic or conservative, may not have estimated deconvolution regime maximum values in cases where we could not find reported $W_{10}$ values.}
\end{deluxetable*}

\startlongtable
\begin{deluxetable*}{CCCCCCCCCCC}
\setlength{\tabcolsep}{0pt}
\tablecaption{Cyclic Merits for LOFAR, NenuFAR, CHIME, \& uGMRT\label{cyclic_merits_2}}
\tablehead{\colhead{Pulsar} & \colhead{\text{LOFAR}$_{200}$} & \colhead{\text{LOFAR}$_{150}$} & \colhead{\text{NenuFAR}$_{48}$} & \colhead{\text{NenuFAR}$_{85}$} & \colhead{\text{CHIME}$_{450}$} & \colhead{\text{uGMRT}$_{150}$} & \colhead{\text{uGMRT}$_{200}$} & \colhead{\text{uGMRT}$_{350}$} & \colhead{Max Freq} & \colhead{Min Freq}}

\startdata
\textrm{B}0031\textrm{$--$}07 &      \textrm{---} &       \textrm{---} &      29.5 (15.9) &      3.67 (1.98) &   123 (66.0) &       661 (356) &      126 (68.1) &       393 (212) &        5 (2) &       3 (1) \\
\textrm{B}0138{+}59 &       121 (110) &       286 (258) &        829 (749) &      97.9 (88.4) &    166 (150) &      1041 (940) &     53.5 (48.3) &       625 (564) &      60 (30) &     34 (17) \\
 \textrm{B}0329{+}54 &     38.7 (19.6) &     92.0 (46.6) &      3526 (1786) &        398 (201) &   11.1 (5.6) &    11104 (5625) &     2414 (1223) &     6682 (3385) &       10 (5) &       5 (2) \\
    \textrm{B}0450{+}55 &       653 (591) &     1524 (1380) &        367 (332) &      45.2 (40.9) &    530 (480) &     1982 (1794) &       263 (238) &     1186 (1074) &      81 (40) &     46 (23) \\
    \textrm{B}0450\textrm{$--$}18 &     \textrm{---} &       \textrm{---} &      1395 (1793) &        274 (352) &    \textrm{---} &     3483 (4478) &       418 (537) &     2015 (2590) &      76 (38) &     41 (20) \\
    \textrm{B}0523{+}11 &     5.89 (2.95) &     13.1 (6.56) &            <0.01 &            <0.01 &  3.54 (1.77) &     31.7 (15.9) &      6.79 (3.4) &     18.7 (9.35) &    273 (136) &    146 (73) \\
    \textrm{B}0525{+}21 &     5.25 (2.64) &     11.4 (5.73) &      10.5 (5.26) &      1.75 (0.88) &  5.19 (2.61) &     16.9 (8.48) &           <0.01 &     9.89 (4.97) &     143 (71) &     74 (37) \\
    \textrm{B}0540{+}23 &     41.1 (20.7) &     88.2 (44.4) &      39.6 (19.9) &      6.93 (3.49) &  14.6 (7.32) &       593 (298) &      129 (64.7) &       346 (174) &    221 (110) &    120 (60) \\
\enddata
\tablecomments{Cyclic merits and upper and lower observing frequency limits on the full deconvolution regime for LOFAR, NenuFAR, CHIME, and uGMRT. Some values that may pass either cyclic merit threshold, optimistic or conservative, may not have estimated deconvolution regime maximum values in cases where we could not find reported $W_{10}$ values.}
\end{deluxetable*}

\bibliography{bibliography}
\bibliographystyle{aasjournalv7}
\end{document}